\documentclass[prd, onecolumn, nofootinbib, floatfix]{revtex4}

\usepackage{epsfig} 
\usepackage{amsmath}
\usepackage{amsbsy,color}

\newcommand{\beq}{\begin{equation}}
\newcommand{\eeq}{\end{equation}}
\newcommand{\beqa}{\begin{eqnarray}}
\newcommand{\eeqa}{\end{eqnarray}}

\def\pa{{\partial}}

\def\ha{\frac{1}{2}}

\def \sn2{\left(S/N\right)^2}

\begin{document}

\title{
Using Cross-Correlations to Calibrate Lensing Source Redshift Distributions \\ {\it Improving Cosmological Constraints
from Upcoming Weak Lensing Surveys}
}
\author{Roland de Putter$^{1}$, Olivier Dor{\'e}$^{1}$, Sudeep Das$^{2}$
\vspace{0.1cm}
}

\affiliation{$^{1}$Jet Propulsion Laboratory, California Institute of Technology, Pasadena, CA 91109\\
\& California Institute of Technology, Pasadena, CA 91125\\
$^{2}$High Energy Physics Division, Argonne National Laboratory, 9700 S Cass Avenue, Lemont, IL 606042} 
\date{\today}

\begin{abstract}

A major challenge for weak gravitational lensing to reach its full potential
as a probe of dark energy is the requirement of sub-percent level calibration of the
source redshift distribution, $dn/dz(z)$. This goal will be difficult to achieve with photometric redshift
information only.
A promising complementary approach is to use cross-correlations between the galaxy number density in the lensing
source sample and that in an overlapping spectroscopic sample to calibrate the source redshift distribution.
In this paper, we study in detail to what extent this cross-correlation method can mitigate
loss of cosmological information in upcoming weak lensing surveys (combined with a CMB prior)
due to lack of knowledge of the source distribution.
We consider a scenario where photometric redshifts are available,
and find that, unless the photometric redshift distribution $p(z_{\rm ph}|z)$
is calibrated very accurately {\it a priori} (bias and scatter known to $\sim 0.002$ for, e.g., EUCLID),
the additional constraint on $p(z_{\rm ph}|z)$ from the cross correlation technique
to a large extent restores the cosmological information originally lost due to the uncertainty in $dn/dz(z)$.
Considering only the gain in photo-$z$ accuracy and not the additional cosmological
information, enhancements of the dark energy figure of merit of up to a factor of 4 (40) can be achieved for
a SuMIRe (Subaru Measurement of Images and Redshifts, the combination of the Hyper Suprime Cam lensing survey
and the Prime Focus Spectrograph redshift survey)-like (EUCLID-like) combination of lensing and redshift surveys.
However, the success of the method is strongly sensitive to our knowledge of
the galaxy bias evolution in the source sample. If this bias is modeled
by a free parameter in each of a large number of redshift bins, we find that a prior
of order $0.01$ is needed on $b^{(p)}_i \sqrt{\Delta z}$ in each redshift slice (where $\Delta z$ is the bin width
and $b^{(p)}_i$ the value of the bias in the $i$-th bin)
to
optimize the gains from the cross-correlation method (i.e.~to approach the cosmology constraints
attainable if the bias were known exactly).
Finally, we study a scenario where no photo-$z$ knowledge is available and
demonstrate that
the redshift distribution itself can be reconstructed very accurately using cross-correlations
(to $\sim 2 \%$ in redshift bins in this particular example) if the galaxy bias is known, but not at all
if the galaxy bias is left free. We also find that,
when a CMB prior is included, 
the ability of weak lensing to constrain cosmological parameters is
mainly sensitive to our knowledge of one particular {\it mode} of $dn/dz(z)$ and that the coefficient of this mode can
in principle be well measured using the cross-correlation approach.

\end{abstract} 

\maketitle 

\section{Introduction}
\label{sec:intro}

Weak gravitational lensing, the subtle distortion of galaxy images by large scale structure along the line of sight,
is a potentially powerful cosmological probe and has been identified as one of the key
future probes of dark energy \cite{detf} (see, e.g.,
\cite{bartschnei01,hoekjain08,obsdereview12} for reviews).
Promising results have already been obtained with existing data,
see for instance \cite{Massetal07,Masseyetal07,Fuetal08,Schrabbetal09,huffetal11,kilbingeretal12,heymansetal13}.
Ongoing and upcoming surveys, such as the Dark Energy Survey (DES, \cite{des}),
the Subaru Hyper Suprime Cam lensing survey (HSC, \cite{hsc}), LSST \cite{lsst} and EUCLID \cite{euclid}, are expected
to deliver shear measurements with sky coverage of more than an order of magnitude larger
than what is currently available and thus have the potential to strongly improve
constraints on dark energy and other cosmological parameters from weak lensing.

However, before the full potential of these upcoming data can be reached,
there are a number of serious challenges that need to be addressed, such
as correcting for the effect of the PSF on galaxy images and
reaching galaxy shape measurements with $< 10^{-3}$ level precision,
modeling non-linear and baryonic effects on the matter power spectrum,
and understanding the contribution to the cosmic shear signal from
intrinsic alignments.

Another main challenge comes from the requirement that the redshift distribution
of lensing source galaxies be known to high precision.
It is this question that motivates the research presented in this paper.
Since the depth and large number density of typical lensing source galaxy samples
preclude the possibility
of obtaining spectroscopic redshifts for all galaxies,
the standard approach is to employ broadband photometry,
using the measured flux through a number of bands to estimate redshifts.
These photometric redshifts (photo-$z$'s) can also be used to divide the sample into tomographic bins,
which allows the extraction of information on redshift evolution of the background cosmology and of large scale structure.

The photo-$z$ estimator can be characterized
by the distribution of estimated redshifts $z_{\rm ph}$, given an object's true redshift $z$,
$p(z_{\rm ph}|z)$.
Since photometric redshifts are essentially based on very low resolution spectra,
they tend to have large uncertainties ($\sigma_z \sim 0.03 - 0.1$). While this large width of the photo-$z$
distribution is not a big problem
for weak lensing studies (the lensing kernel is very broad anyway), the shape of the distribution needs to be known
to high precision to avoid biasing cosmological parameter estimates.
For example, \cite{mahuhut06} (see also \cite{hut02,hutetal04,hutetal06}) have shown that both the width and the bias of this distribution
need to be known to better than about $0.003 - 0.01$ (depending on the dark energy model
considered) for future surveys.

In order to characterize the photo-$z$ distribution, spectroscopic redshifts are required for a large, representative subsample of
galaxies\footnote{Alternatively, a large spectroscopic sample could even
be used to directly characterize the source redshift distribution without using photo-$z$'s.
However, there would be no way to do tomography in that approach.}
(in addition, depending on the photo-$z$ method used,
spectroscopic redshifts are needed to ``train'' the photo-$z$ estimator).
For upcoming surveys, this would require samples of $\sim 10^5$ faint ($i \sim 22 - 26$) spectroscopic galaxies
with large redshift success rates and few redshift failures (e.g.~\cite{cunhaetal12}).
For many upcoming weak lensing surveys,
it is not at all clear if appropriate spectroscopic samples will be available.

A complementary/alternative method that can be used either to directly estimate the redshift distribution
of a (source) galaxy sample, or to calibrate the photo-$z$ distribution,
was proposed in \cite{newman08}.
In this  method, angular cross-correlations between the number density of source galaxies
and the number density of an overlapping sample of spectroscopic galaxies in various redshift bins
are used to deduce the average source galaxy number density in these bins\footnote{This is but one of the ways in which
complementarity between (overlapping) imaging and spectroscopic surveys can be exploited. For example,
\cite{caibern12,gaztaetal12} studied the expected gains from using the full cosmological information encoded in both
the weak lensing and galaxy clustering signal.}.
The spectroscopic galaxies are only required to cover the same volume (or a subvolume) as the source galaxies so that
they trace the same matter density modes and are {\it not} required to be drawn from the same sample
as the source galaxies.

While the focus of this article is the application to weak lensing source samples, the cross-correlation method can be applied
more generally to any sample for which the redshift distribution is desired.
The expected performance of the cross-correlation method has been studied in detail in previous works
\cite{newman08,mathnewman10,schulz10,mattnewman12,mcwhite13}
(see also \cite{schneideretal06,bernhut10})
and interesting results have been obtained on how the success of the method depends on survey properties,
and on what are the main potential obstacles. While conclusions vary somewhat between works depending on the focus,
the cross-correlation technique looks promising based on these studies.
The approach has also been applied to existing data with some success \cite{phillipps85,masjedietal06,hoetal08,menardetal13}.

The goal of this paper is to quantify to what extent measurements of the lensing source distribution via
the cross-correlation method improve the expected cosmological constraints from cosmic shear data.
In other words, we wish to know to what degree the cross-correlation method mitigates the loss of information
in a cosmic shear analysis due to uncertainty in the source distribution. This question has only indirectly been studied
in previous work. Typically, the accuracy of redshift distribution reconstruction is ascertained,
translated to an uncertainty on the average redshift and the redshift scatter/width of a sample,
and then compared to requirements on these quantities from weak lensing forecasts available elsewhere in the literature (e.g.~\cite{mahuhut06}).
While these previous studies show that the cross-correlation method is promising, a more quantitative study
would provide more insight. In this paper, we therefore present an integrated analysis of
the use of the cross-correlation technique in conjunction with a forecast of cosmology constraints from cosmic shear,
explicitly showing how dark energy (and other) constraints from cosmic shear are affected by the information
on the lensing source redshift distribution obtained from the cross-correlations.

We will consider two examples of upcoming combinations of overlapping lensing and redshift surveys:
(1) SuMIRe, the combination of the HSC lensing survey and the PFS spectroscopic galaxy survey \cite{pfsreport12},
and (2) EUCLID, which will carry out both types of surveys.
We will use the Fisher matrix formalism to forecast parameter constraints, with a focus on the dark energy equation of state.
The resulting uncertainties approximately correspond to the uncertainties expected to be obtained from a maximum likelihood estimator
or optimal quadratic estimator (\cite{mcwhite13}).

One of the main potential problems with the cross-correlation method is that
the effect of the source redshift distribution on the observed angular cross-correlations
is degenerate with redshift evolution of the source galaxy bias
(see also \cite{schulz10,bernhut10,mathnewman10,mcwhite13}).
This means that the strength of the cross-correlation technique
crucially depends on the (prior) knowledge of this bias evolution.
We will in this work go significantly beyond previous studies of this issue
by allowing for an arbitrary redshift dependence of the galaxy bias (defined in narrow redshift slices)
and studying how the source redshift distribution reconstruction and cosmological constraints
depend on the priors places on the bias evolution.

We consider two scenarios for the application of the cross-correlation method, roughly dividing the paper
into two parts.
In the first part of the paper (Sections \ref{sec:shear th} - \ref{sec:cc results}),
we will study the use of the cross-correlation technique in combination
with photometric redshift information.
This analysis follows the more or less standard method in the literature,
where the source distribution in tomographic bins is determined by the shape
of $p(z_{\rm ph}|z)$, taken for simplicity to be a Gaussian.
The information from cross-correlations between the source galaxy number density and spectroscopic galaxies
then comes in as a way to measure the parameters (the bias and width specified at different redshift) defining $p(z_{\rm ph}|z)$.
Questions of particular interest we will address, in addition to that of the degeneracy with bias evolution,
are the dependence on the smallest scale used ($\sim k_{\rm max}^{-1}$) in the cross-correlation analysis,
and the dependence on how well the photo-$z$ distribution was calibrated (e.g.~by using a deep spectroscopic galaxy sample)
before the information from cross-correlations is employed.

The outline of this first part of the paper is as follows.
In Section \ref{sec:shear th}, we review the relevant expressions
describing the cosmic shear signal and the roles of the source redshift distribution
and the photo-$z$ parameters.
In Section \ref{sec:wlsurvey},
we briefly describe the assumed survey specifications
for the HSC and EUCLID lensing surveys.
We then forecast cosmological constraints from these surveys (including a CMB prior)
in Section \ref{sec:results wl}, and highlight the dependence on the assumed knowledge of the source
distribution.
Next, in Section \ref{sec:cc th}, we review the formalism of the cross-correlation technique
and describe our galaxy bias priors,
while in Section \ref{sec:spec surveys} we describe the PFS and EUCLID
spectroscopic surveys. The main results of the first part of the paper are then presented
in Section \ref{sec:cc results}, where we consider dark energy (and other) constraints
when combining the information on the source redshift distribution from the cross-correlations
with the cosmological information encoded in the weak lensing power spectra.

In the (much shorter) second part of the paper, Section \ref{direct}, we consider a different approach
and study the case of an unknown source distribution without photo-$z$ information.
On the one hand, we will quantify how well such a distribution can be measured in narrow redshift bins directly
from the cross-correlations with the spectroscopic sample (this is also what has been done in previous works).
On the other hand, we will ask which components of the source redshift distribution need to be known, and how well,
in order for cosmic shear constraints not to be impacted. We then compare these two results to gain more insight into
how the cross-correlation method improves cosmological constraints from weak lensing
and therefore into the results of the prior sections.

We end the article with a Discussion and Conclusions in Section \ref{sec:disc}.

\section{Cosmic Shear (Theory)}
\label{sec:shear th}

The weak lensing convergence of source galaxies in a tomographic bin $i$ is given by
\beq
\kappa_i(\hat{n}) = \int dz \, W_i(z) \, \delta(D(z) \, \hat{n}, z),
\eeq
where $D(z)$ the radial coordinate distance to redshift $z$,
$\delta(\vec{x}, z)$
is the relative matter overdensity, and the kernel
\beq
\label{eq:kernel}
W_i(z) = \frac{3}{2} H_0^2 \, \Omega_m \, (1 + z) \, D(z) \, H^{-1}(z) \, \int dz_S \, f_i(z_S) \, \frac{D(z, z_S)}{D(z_S)}.
\eeq
Here, $H_0$ and $\Omega_m$ are the present values of the Hubble rate and the matter density relative to critical
and $D(z, z')$ is the radial coordinate distance from
redshift $z$ to redshift $z'$ (in a spatially flat universe, $D(z, z') = D(z') - D(z)$.
Finally, $f_i(z)$ is the redshift distribution of source galaxies in the $i$-th bin, normalized to unity,
i.e.
\beq
\label{eq:fz}
f_i(z) \equiv \left( \int dz' \, \frac{d n_{i}}{dz}(z') \right)^{-1} \, \frac{d n_{i}}{dz}(z),
\eeq
with $d n_{i}/dz(z)$ the number of galaxies per steradian per unit redshift.
We summarize these properties in the left column of Table \ref{tab:wlspecs}.

The power- and cross-spectra of the convergence are (in the Limber approximation \cite{Limber:54}) given by
\beq
\label{eq:kappa spec}
C_l^{ij} = \int dz \, \frac{H(z)}{D^2(z)} \, W_i(z) \, W_j(z) \, P\left(\frac{\ell + \ha}{D(z)}, z\right),
\eeq
where $P(k, z)$ is the matter power spectrum at redshift $z$.
The convergence spectra probe cosmology both through their dependence on the matter power spectrum
and through the dependence on the distances and expansion rate appearing in the line of sight integral.
An important advantage of gravitational lensing over many other probes of large scale structure is that it
directly measures the matter density field as opposed to a tracer of it, avoiding the need to model the bias of
such a tracer (note, however, that Eq.~(\ref{eq:kappa spec}) assumed general relativity to relate metric perturbations
to the matter density).

To describe the source redshift distribution, we use what is more or less the standard in the literature for forecasts
that include photometric redshift calibration uncertainties, see e.g.~\cite{mahuhut06,hutetal06,mabernstein08,hearinetal10}.
We assume that the galaxies have photometric redshifts $z_{\rm ph}$,
characterized by a distribution $p(z_{\rm ph}|z)$ (the probability density for the photometric redshift, given the true galaxy redshift)
and that tomographic bins are defined by cuts in $z_{\rm ph}$. The (true) redshift distribution in a bin is then
\beq
\label{eq:bindist}
\frac{dn_i}{dz}(z) = \frac{dn}{dz}(z) \, \int_{z_i^{\rm low}}^{z_i^{\rm high}} dz_{\rm ph} \, p(z_{\rm ph}|z),
\eeq
where $z_i^{\rm low}$ and $z_i^{\rm high}$ are the boundaries of the bin. The true redshift distribution of the full sample,
$\frac{dn}{dz}(z)$, is assumed to be known and will depend on the survey under consideration (given explicitly in Section \ref{sec:wlsurvey}).
The normalized distribution in a given bin can be trivially obtained from Equations (\ref{eq:bindist})
and (\ref{eq:fz}).

As a baseline model for the photo-$z$ distribution, we assume a Gaussian
\beq
\label{eq:phzdist}
p(z_{\rm ph}|z) = \frac{1}{\sqrt{2 \pi} \sigma_z(z)} \, e^{-\ha (z_{\rm ph} - z - b_z(z))/\sigma_z^2(z)},
\eeq
characterized by a scatter $\sigma_z(z)$ and a bias $b_z(z)$.
We parametrize the photo-$z$ scatter and bias by the values of $\sigma_z$ and $b_z$ at 11 equally spaced redshifts
in the range $z = 0 - 3$. The values of $\sigma_z(z)$ and $b_z(z)$ at arbitrary redshift are then obtained by interpolation.
We assume a fiducial of $\sigma_z(z) = 0.05 (1 + z)$ and $b_z(z) = 0$. To incorporate uncertainty in the photo-$z$ distribution,
the 11 pairs of $(\sigma_z, b_z)$ values are treated as free parameters, on which priors can later be imposed.
In reality, the photo-$z$ distribution is typically not Gaussian, but the Gaussian form serves as a simple ansatz
with which to study uncertainty in the width and average of the photo-$z$ distribution.
For future work, it would be interesting to include, e.g., skewness of the distribution,
and the possibility of outliers to study catastrophic redshift failures.

The model described above clearly presents an oversimplified description of the use of photometric redshifts with a real survey,
but should suffice for a first investigation
of how useful cross-correlations are for optimizing cosmological information in cosmic shear by calibrating the source
redshift distribution.
Another simplification we will make is to ignore the effect of intrinsic alignments (e.g.~\cite{penleesel00,catelanetal01,hirsel04,heymansetal13,heavetal00}),
optimistically assuming that for the galaxy types where this effect matters, it can be modeled and removed.

\section{Weak Lensing Surveys}
\label{sec:wlsurvey}

\begin{table*}[hbt!]
\begin{center}
\small
\begin{tabular}{c|cc}
\hline\hline
$\,$ & HSC WL survey & EUCLID WL survey \\
\hline
 number density & $\bar{n}_A = 20 $ arcmin$^{-2}$ & $\bar{n}_A = 30 $ arcmin$^{-2}$ \\
 sky coverage  & $1500$ deg$^2$ & $15000$ deg$^2$ \\
 shape noise  & $\sigma(\gamma) = 0.22$ & $\sigma(\gamma) = 0.22$ \\
 redshift distribution & $dn/dz \propto z^2 e^{-z/z_0}, \, \langle z \rangle = 1.0 \quad $ & $dn/dz \propto z^2 e^{-(z/z_0)^{3/2}}, \, \langle z \rangle = 0.96 \quad $ (median $0.9$) \\
 tomography & 3 bins & 6 bins \\
 multipole range & $\ell = 20 - 2000$ & $\ell = 20 - 2000$ \\
\hline\hline
\end{tabular}
\caption{Properties of the weak lensing surveys considered in this article.
}
\label{tab:wlspecs}
\end{center}
\end{table*}

\begin{figure*}
  \begin{center}{
  \includegraphics[width=0.48\columnwidth]{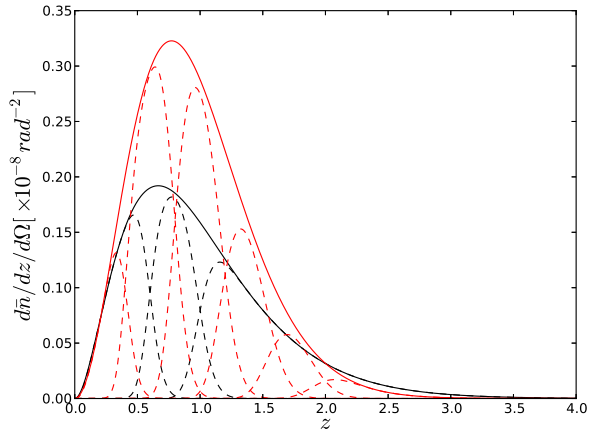}
  }
  \end{center}
  \caption{Source distribution of full sample (solid) and of galaxies in tomographic bins (dashed) for HSC (black)
  and EUCLID (red). For HSC, the source bins are defined by $z_{\rm photo} = 0.0 - 0.6 - 1.0 - 4.0$
  and for EUCLID, they are defined by the cuts $z_{\rm ph} = 0.0 - 0.4 - 0.8 - 1.2 - 1.6 - 2.0 - 3.5$.
  }
  \label{fig:sourcedist}
\end{figure*}

We consider two upcoming weak lensing surveys: the Subaru Hyper Suprime-Cam (HSC) wide field survey \cite{hsc}, starting in 2013,
and
that of the EUCLID satellite \cite{euclid}, scheduled to launch in 2020. The main specifications of each experiment are listed in Table \ref{tab:wlspecs}.
For HSC, they are based on \cite{OguriTakada11} and for EUCLID on \cite{amendolaetal12}.
Figure \ref{fig:sourcedist} shows the source distributions for the full sample and in the individual bins for each experiment.

\section{Effect of uncertainty in photo-z distribution}
\label{sec:results wl}

We use the Fisher matrix formalism (see, e.g., \cite{TegTayHeav97}) to forecast constraints from weak lensing on cosmological (and photo-$z$)
parameters.
The matter power spectra that serve as input for this calculation are computed using the public Boltzmann code CAMB \cite{LewChalLas00}.
We consider a spatially flat universe with dynamical dark energy parametrized by the equation of state
function\footnote{The varying dark energy equation of state is implemented using the parametrized post-Friedmann (PPF)
description \cite{ppf}.}
$w(a) = w_0 + w_a (1 - a)$. We thus have eight parameters in total, with fiducial values
$\omega_b = 0.02258, \omega_c = 0.1109, \Omega_\Lambda = 0.734, \sigma_8 = 0.8, n_s = 0.96, w_0 = -1, w_a = 0$ and
$\tau = 0.086$, where $\tau$ is the optical depth to reionization.
Weak lensing on its own places only weak constraints on the dark energy parameters because of parameter degeneracies.
Hence, we combine weak lensing with a forecasted CMB prior from the Planck \cite{planckoverview,planckpowlik,planckcosmoparam}
experiment. The Planck Fisher matrix employed here
was calculated using the specifications given in Table \ref{tab:specs}. We assume only multipoles $\ell < 2000$ are used
and neglect CMB lensing to be on the conservative side and to avoid having to model covariance between the CMB spectra and
the lensing spectra. Forecasted constraints much stronger than the ones we will show could be obtained by adding more data sets,
but since we would like to isolate as much as possible the constraining power of weak lensing, we will not follow this path.
We study the results for different surveys below. While we will consider uncertainties on all parameters,
the main quantity of interest will be the dark energy figure of merit (FOM, \cite{detf}), defined here as
\beq
{\rm FOM} = \left( {\rm Det}({\rm Cov}[w_0,w_a]) \right)^{-\ha},
\eeq
which is inversely proportional to the area enclosed by a fixed confidence level contour in the $w_0 - w_a$ plane.
Note that, unlike in the definition of the Dark Energy Task Force, we do not marginalize over spatial curvature $\Omega_K$,
but instead fix it to zero.

\begin{table*}[hbt!]
\begin{center}
\small
\begin{tabular}{c|ccccc}
\hline\hline
$\,$ & $\nu$ & $\theta_{\rm FWHM}$ & $\Delta_T[\mu K$ - arcmin$]$ & $\Delta_P[\mu K$ - arcmin$]$ & $f_{\rm sky}$ \\
\hline
 Planck & 100 GHz & 9.5' & 80 & 114 & 0.8 \\
  $\,$  & 143 Ghz & 7.1' & 46 & 77 & 0.8 \\
  $\,$  & 217 Ghz & 4.7' & 70 & 122 & 0.8 \\
\hline\hline
\end{tabular}
\caption{Planck specifications used for Fisher forecasts. We use $\ell_{\rm max} = 2000$ both for temperature and polarization
and neglect information from CMB lensing.
}
\label{tab:specs}
\end{center}
\end{table*}

We use linear power spectra as input to calculate the derivatives going into the Fisher matrix
and also to compute the covariance matrix of the observables. The latter calculation assumes Gaussianity
of the shear field, leading to a diagonal covariance matrix. As we briefly illustrate at the end of Section \ref{subsec:hsc shear},
using the information in the non-linear power spectra would lead to significantly stronger forecasted constraints.
However, using the non-linear signal, but ignoring the non-Gaussian contributions (specifically the off-diagonal contributions)
to the covariance matrix, overestimates the constraining power of weak lensing (see, e.g., \cite{takjain09,kiesslingetal11}).
While this mainly manifests itself in an underestimate of the multi-dimensional volume of the allowed region
in parameter space and individual parameter uncertainties are not affected strongly, we still prefer
to present conservative constraints that do not use the information in the non-linear regime at all.
Since our main interest in this work is in the {\it dependence} of parameter constraints on the
level of knowledge of the source redshift distribution, rather than in the exact values of the forecasted
uncertainties or FOM, this is not a choice of great consequence.

\subsection{HSC}
\label{subsec:hsc shear}

\begin{figure*}
  \begin{center}{
  \includegraphics[width=0.48\columnwidth]{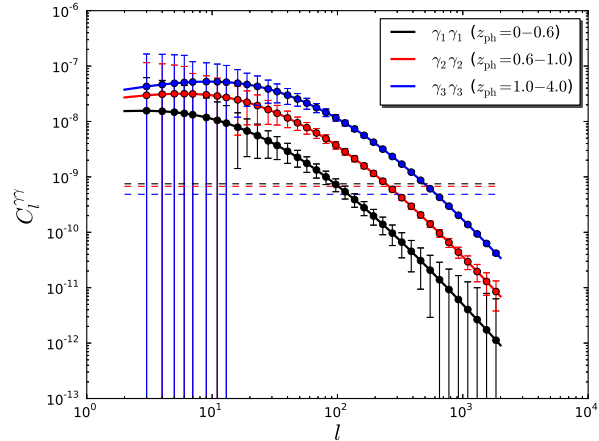}
  }
  \end{center}
  \caption{Shear angular power spectra in three tomographic bins for the HSC lensing survey assuming our fiducial cosmology (solid).
  The dashed horizontal lines indicate the shape noise power spectra.
  }
  \label{fig:clshear}
\end{figure*}

We show the cosmic shear angular power spectra (solid) in our fiducial cosmology, the shape noise power
spectra (dashed) and the uncertainty on the binned angular power spectrum in Figure \ref{fig:clshear}.
A comparison of the noise and signal spectra shows that the measurement becomes noise dominated above a
critical multipole in the range $\ell = 100 - 1000$ depending on the redshift bin (and on the bin widths chosen).
However, by averaging
over the large number of available modes, the power spectrum itself can be measured with high accuracy to much higher multipoles,
as is shown by the error bars.
Using these spectra and their derivatives
with respect to cosmological and photo-$z$ parameters, we construct a Fisher matrix.

In Table \ref{tab:DDE}, we show the resulting parameter uncertainties and dark energy figure of merit.
When perfect knowledge of the photo-$z$ parameters is assumed (i.e.~fixing the $\sigma_z$ and $b_z$ parameters), adding weak lensing to Planck
significantly improves cosmological constraints, causing an increase of the dark energy FOM
by a factor 20 and tightening parameter uncertainties by up to a factor of five.
The third column, however, shows the results in the case where the 22 photo-$z$ parameters are left free. In this case,
these parameters are self-calibrated by the cosmic shear data. As can be expected, there is so much freedom in the photo-$z$ parameter space,
that the weak lensing spectra leave them essentially unconstrained. In other words, the self-calibration
is ineffective. As a result, there is a strong degradation of cosmological information, with the
parameter uncertainties and dark energy figure of merit returning to their values in the case of Planck only. Thus, when no knowledge
of the photo-$z$ parameters is assumed, weak lensing does not add any cosmological information compared to the CMB.

In reality, one will have some more knowledge about the photo-$z$ distribution, coming from our understanding of the photo-$z$ estimator
and its calibration with spectroscopic galaxies. This knowledge can in our simplified model be captured
by an external prior on the photo-$z$ parameters. The question of what prior level is needed to obtain optimal cosmological
constraints has been studied in great detail in \cite{mahuhut06,hutetal06}. We find here that if we place a prior on each
photo-$z$ parameter of $\sigma(\sigma_z) = \sigma(b_z) = 0.001 (0.01)$, we recover all - most of the cosmological information
(FOM $ = 9.5 (7.0)$).
This is consistent with the findings in the above mentioned works.
In Sections \ref{sec:cc th} - \ref{sec:cc results} we will show to what extent cross correlations between the source galaxies
and an overlapping sample of spectroscopic galaxies can calibrate the photo-$z$ parameters and recover the cosmological
information in weak lensing. We will there also include the effect of having prior knowledge of the photo-$z$ parameters.

Finally, we show in Table \ref{tab:DDE 2} the forecasted parameter constraints found when using non-linear matter power spectra
(both for the derivatives and the covariances) to calculate the Fisher matrix. Using the non-linear information
improves the dark energy figure of merit by a factor of three in the case of known photo-$z$ parameters.
Moreover, the non-linear power spectrum helps break some of the degeneracy between cosmological parameters and
the photo-$z$ distribution (as discussed in \cite{hearinetal12}), as even without a prior on the photo-$z$ parameters, we now find that lensing (slightly)
improves constraints relative to the Planck-only case. For reasons discussed above, we will from now on
only use the linear power spectra in our Fisher matrix calculations, but it is good to keep in mind that
this is gives rather conservative error estimates.

\begin{table*}[hbt!]
\begin{center}
\small
\begin{tabular}{c|ccc}
\hline\hline
$\sigma(p)$ &  Planck & Planck + $\gamma \gamma$ (known $\sigma_z, b_z$) & Planck + $\gamma \gamma$ (``free'' $\sigma_z, b_z$) \\
\hline\hline
$\omega_b$ & 0.00013 & 0.00011  & 0.00013 \\
$\omega_c$ & 0.0011 & 0.00069 & 0.0011 \\
$\Omega_\Lambda$ & 0.18 & 0.051 & 0.17 \\
$n_s$ & 0.0033 & 0.0027 & 0.0033 \\
$\sigma_8$ & 0.20  & 0.043 & 0.18 \\
$w_0$ & 1.5 & 0.61 & 1.5 \\
$w_a$ & 3.7 & 1.6 & 3.7 \\
FOM$=1/\sqrt{{\rm Det Cov}}$ & 0.47 & 9.5 & 0.52 \\
\hline\hline
\end{tabular}
\caption{Forecasted cosmological constraints for the HSC weak lensing survey.
Uncertainties and dark energy figure of merit are shown for Planck (left column),
Planck + cosmic shear with known/fixed photo-$z$ parameters (middle), and
Planck + cosmic shear with {\it a priori} unknown photo-$z$ parameters (right).
The photo-$z$ scatter $\sigma_z(z)$ and bias $b_z(z)$ are specified at 11 redshifts
in the range $z = 0-3$ and interpolated in between.
}
\label{tab:DDE}
\end{center}
\end{table*}

\begin{table*}[hbt!]
\begin{center}
\small
\begin{tabular}{c|ccc}
\hline\hline
$\sigma(p)$ &  Planck & Planck + $\gamma \gamma$ (known $\sigma_z, b_z$) & Planck + $\gamma \gamma$ (``free'' $\sigma_z, b_z$) \\
\hline\hline
$\omega_b$ & 0.00013 & 0.000097 & 0.0012 \\
$\omega_c$ & 0.0011 & 0.00044 &  0.0011 \\
$\Omega_\Lambda$ & 0.18 & 0.018 & 0.13 \\
$n_s$ & 0.0033 & 0.0023 & 0.0031 \\
$\sigma_8$ & 0.20  & 0.019 & 0.15 \\
$w_0$ & 1.5 & 0.29 & 0.85 \\
$w_a$ & 3.7 & 0.82 & 1.6 \\
FOM$=1/\sqrt{{\rm Det Cov}}$ & 0.47 & 29 & 1.7 \\
\hline\hline
\end{tabular}
\caption{Same as Table \ref{tab:DDE}, but with non-linear shear spectra instead.}
\label{tab:DDE 2}
\end{center}
\end{table*}

\subsection{EUCLID}
\label{subsec:euclid gg}

\begin{figure*}
  \begin{center}{
  \includegraphics[width=0.48\columnwidth]{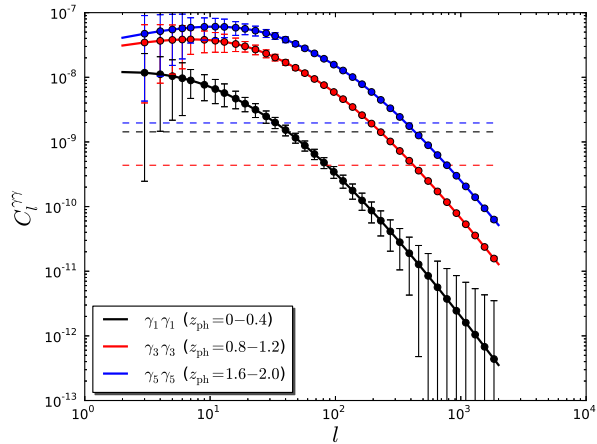}
  }
  \end{center}
  \caption{Shear angular power spectra in three tomographic bins for the lensing component
  of the EUCLID survey assuming our fiducial cosmology (solid).
  The dashed horizontal lines indicate the shape noise power spectra.
  }
  \label{fig:clshear euclid}
\end{figure*}

For EUCLID's lensing survey, the angular power spectra and noise spectra are shown in
Figure \ref{fig:clshear euclid} for a subset of the six tomographic bins, showing that the shear measurements
become noise dominated at scales $\ell > \ell_{\rm max}$, with $\ell_{\rm max} = 30 - 600$, depending on the bin.
The error bars on the binned angular spectra are significantly smaller than in the case of HSC,
mainly because the ten times larger sky coverage for EUCLID
(note, however that because of the different binning choices, one cannot make a direct quantitative comparison
between the two figures).

The forecasted parameter uncertainties are given in Table \ref{tab:gg euclid}. As in the case of HSC, cosmic shear
strongly improves cosmological parameter constraints provided that the photo-$z$ distribution is known. With EUCLID,
the improvement is even more spectacular than before, giving more than a factor 300 increase in FOM.
Allowing freedom in the photo-$z$ parameters (without an external prior) degrades this information again,
although the constraints are still slightly better than with CMB only. The requirement on an external prior
on the photo-$z$ parameters is a bit more stringent than before, with priors $\sigma(\sigma_z) = \sigma(b_z) = 0.001 - 0.01$
giving figures of merit FOM $ = 150 - 26$, so that again subpercent level priors are required to fully exploit the power of
weak lensing.

\begin{table*}[hbt!]
\begin{center}
\small
\begin{tabular}{c|ccc}
\hline\hline
$\sigma(p)$ &  Planck & Planck + $\gamma \gamma$ (known $\sigma_z, b_z$) & Planck + $\gamma \gamma$ (``free'' $\sigma_z, b_z$) \\
\hline\hline
$\omega_b$ & 0.00013 & 0.000093  & 0.00013 \\
$\omega_c$ & 0.0011 & 0.00031 & 0.0011 \\
$\Omega_\Lambda$ & 0.18 & 0.013 & 0.13 \\
$n_s$ & 0.0033 & 0.0021 & 0.0031 \\
$\sigma_8$ & 0.20  & 0.011 & 0.12 \\
$w_0$ & 1.5 & 0.14 & 1.4 \\
$w_a$ & 3.7 & 0.36 & 3.6 \\
FOM$=1/\sqrt{{\rm Det Cov}}$ & 0.47 & 162 & 1.5 \\
\hline\hline
\end{tabular}
\caption{Forecasted cosmological constraints for the EUCLID weak lensing survey.
Uncertainties and dark energy figure of merit are shown for Planck (left column),
Planck + cosmic shear with known/fixed photo-$z$ parameters (middle), and
Planck + cosmic shear with {\it a priori} unknown photo-$z$ parameters (right).
The photo-$z$ scatter $\sigma_z(z)$ and bias $b_z(z)$ are specified at 11 redshifts
in the range $z = 0-3$ and interpolated in between.
}
\label{tab:gg euclid}
\end{center}
\end{table*}

\section{Cross-Correlations (Theory)}
\label{sec:cc th}

We now consider including a galaxy sample with spectroscopic redshifts that covers the same area of the sky
as the photometric sample of lensing source galaxies.
We stress that the galaxy selection of this sample does not need to overlap with that of the lensing source galaxy sample.
All that matters is that the two samples overlap in surveyed volume, so that the galaxy densities
trace the same underlying dark matter density. In practice, the spectroscopic sample will most likely
consist of more luminous galaxies, with a smaller number density, than the lensing source sample.
Allowing for the possibility of dividing both the photometric (label $p$)
and spectroscopic (label $s$) samples into bins,
the auto- and cross-correlations of the overdensities in these bins are given by
Equation (\ref{eq:kappa spec}),
\beq
\label{eq:crosscl}
C_l^{ij} = \int dz \, \frac{H(z)}{D^2(z)} \, W_i(z) \, W_j(z) \, P\left(\frac{\ell + \ha}{D(z)}, z\right) \nonumber,
\eeq
where now the kernels are given by
\beq
W^i(z) = b^{(p)}(z) \, f_i(z)
\eeq
for the bin(s) of photometric source galaxies, where $b^{(p)}(z)$ is the galaxy bias of this sample
and $f_i(z)$ the normalized redshift distribution,
and by
\beq
W^i(z) = b^{(s)}(z) \, f^{(s)}_i(z)
\eeq
for the spectroscopic bins, with $b^{(s)}(z)$ and $f^{(s)}_i(z)$
the bias and distribution of the spectroscopic galaxies.
Equation (\ref{eq:crosscl}) assumes the Limber approximation, which, for the auto-spectra, is appropriate
for scales $\ell \gtrsim D/\Delta D$ (see \cite{loveafsh08}), where $D$ and $\Delta D$
are the distance to, and width of the redshift slice.
\cite{mcwhite13} have recently shown that the Limber approximation is appropriate for the application considered here
because most of the information on the source redshift distribution comes from scales well into the $\ell > D/\Delta D$ regime.
To be careful, we choose $\ell_{\rm min} = 20$ in our forecast so that the above inequality is satisfied approximately
for all included modes (the largest/{\it worst case} value of $D/\Delta D$ will actually be $35$, but, again,
not much information comes from the modes at $\ell < D/\Delta D$, so our choice of $\ell_{\rm max}$ should suffice for a forecast).

We will divide the spectroscopic sample into a large number of narrow redshift slices and will treat the spectroscopic
redshifts as infinitely accurate so that the redshift distribution within a slice has zero weight outside of its defining redshift bounds.
In the limit of an infinitely narrow bin at redshift $z_i$, $f^{(s)}_i(z) \to \delta^{(D)}(z - z_i)$ (with $\delta^{(D)}$ the Dirac
delta function),
so that the cross correlation with a photometric bin $j$ becomes
\beq
\label{eq:sp}
C_l^{ij} = \frac{H(z_i)}{D^2(z_i)} \, b^{(s)}(z_i) \, b^{(p)}(z_i) \, f_j(z_i) \, P\left(\frac{\ell + \ha}{D(z_i)}, z_i \right)
\eeq
(note that the Limber approximation can still be applied here because the photometric galaxy distribution is assumed to
be spread out in redshift).
The cross-correlation is thus proportional to, $f_j(z_i)$, the redshift distribution of photometric galaxies in the $j$-th source bin at redshift $z_i$.
This is what motivates cross-correlating with a large number of spectroscopic redshift bins
to reconstruct the full function $f_j(z)$.
The spectroscopic galaxy bias in Equation (\ref{eq:sp}) can in principle be obtained from the auto-spectrum of the
spectroscopic galaxies. Moreover, the auto-spectrum of the photometric sample may contain additional information on
$f_j(z)$ as well. We therefore include in our analysis not just the photo-spec cross correlations ($ps$), but also
the auto-correlations ($pp$ and $ss$).

We will neglect the effect of magnification bias on the auto- and cross-spectra. Magnification bias may
act as a double-edged sword (see also \cite{mcwhite13} for an interesting discussion of magnification bias on
redshift distribution estimation from cross-correlations). On the one hand, it introduces a correction to Equation (\ref{eq:sp}),
so that the cross-spectra are no longer directly proportional to $f_j(z_i)$ and the correction is non-trivial
to model because of the uncertainty in the power law index of the source galaxy number vs.~flux threshold relation,
$\alpha^{(p)}$. On the other hand, the additional signal may help break the bias degeneracy we will discuss below.
We leave further investigation of this possibility for future work.

Of course, to extract {\it all} information from the shear/convergence field and the galaxy overdensities,
one would use all possible correlations, including for example the cross correlations between shear and
the spectroscopic galaxies (galaxy-galaxy lensing). However, we here wish to focus specifically on the use
of the spectroscopic galaxies to measure the lensing source distribution. For this reason,
we will only consider the $sp$, $ss$ and $pp$ spectra (in addition to the lensing power spectrum discussed in the previous section
and a CMB prior). Moreover, the galaxy overdensities do not only carry information on the redshift distribution, but also
direct cosmological information. For the reason explained above however, and to be on the conservative side,
we will first calculate a Fisher matrix for the parameters determining $f_j(z)$, marginalized over the cosmological parameters,
using $sp$, $ss$ and $pp$.
We then add this Fisher matrix to the full Fisher matrix from weak lensing and CMB. This way, we are not including directly any cosmological
information encoded in the galaxy clustering.
Moreover, any degeneracy between the effect of the source redshift distribution and the effect of cosmological parameters
is thus explicitly marginalized over, unlike in previous studies.

\subsection{The Role of Galaxy Bias Evolution}
\label{subsec:biasevol}

Equation (\ref{eq:kappa spec}) shows that all the spectra involving the photometric sample are only sensitive to
the bias and the redshift distribution through their product, $b^{(p)}(z) \, f_j(z)$, giving rise to an exact degeneracy
between the two functions \cite{schulz10,bernhut10,mathnewman10,mcwhite13}.
This is a potentially serious challenge for the cross-correlation technique. In the following, we will first show how well this technique
works in the case where the photometric galaxy bias function is known exactly {\it a priori}.
We will also study the more realistic case where it is not, and we ask what prior is needed on the galaxy bias for the method to still
be useful.

To do this, we will treat the galaxy bias of the source sample as scale independent (appropriate in the linear regime),
with redshift evolution modeled as a piecewise constant function in redshift bins, with the value in each bin
given by a parameter $b_i^{(p)}$. We assume a fiducial $b_i^{(p)} = 1$ for all $i$.
The binning choice will be discussed for each survey in Section \ref{sec:spec surveys}.
The case of {\it a priori} unknown galaxy bias is reproduced by leaving these parameters free,
without an external prior. We will then consider two types of bias priors:

\begin{itemize}
\item
Imposing independent priors $\sigma(b^{(p)}_i)$ on the binned bias parameters.
In this case, the required prior depends on the choice of redshift bins in which the galaxy
bias is assumed piecewise constant. Specifically, in the limit of a large number of bins (bin width $\Delta z$ small),
the scaling is approximately $\sigma(b^{(p)}_i)|_{\rm req.} \propto (\Delta z)^{-\ha}$. To quantify the prior in a binning-independent
manner, we thus define
\beq
\sigma(b^{(p)}_i) = \sigma_{\rm diag}^{\rm bias}/\sqrt{\Delta z}
\eeq
for all redshift bins $i$ and will quote the quantity $\sigma_{\rm diag}^{\rm bias}$.
$\sigma^{\rm bias}_{\rm diag}$ can be thought of as the prior on the average galaxy bias
in a bin of fixed width $\Delta z = 1$.

\item

Assuming the redshift dependence of the bias to be linear in redshift (expanded around a central redshift $z_0 = 1.0$, the precise
choice of which is irrelevant),
\beq
b^{(p)}_i = b_0^{(p)} + b'^{(p)}  \, (z_i - z_0),
\eeq
and applying a prior to the coefficient,
\beq
\sigma(b'^{(p)}) \equiv \sigma_{\rm lin}^{\rm bias}.
\eeq
Here, $z_i$ is the central redshift of the $i$-th galaxy bias
redshift bin. We note that the prior on any redshift independent contribution to $b^{(p)}(z)$
is irrelevant, as a constant galaxy bias is not degenerate with $f(z)$ because of the normalization constraint
on the latter function.

\end{itemize}

We will not study the important question of {\it how} to obtain a prior on the bias evolution of the source sample.
This is a difficult questions, deserving of a paper of its own.
As stated above, the source for the number density per unit area of the source sample, to linear order,
is directly proportional to the product $b^{(p)}(z) f_j(z)$ (considering the $j$-th tomographic bin)
so that this quantity alone can never be used to break the degeneracy. However, as we discussed briefly,
there is also a magnification bias contribution. The source term for this contribution is similar in nature to
the source for the cosmic shear signal itself. It thus does not depend on galaxy bias and in principle carries information
on the source distribution that is independent of galaxy bias. While this information is unfortunately not localized in redshift,
it has been found that it may still be possible to use it to constrain the galaxy bias to $\sim 10 \%$ \cite{mcwhite13}.
Another way of avoiding the bias degeneracy is to consider the signal in the non-linear regime, where
the one-halo term and non-linear bias may carry additional information. Finally, it may be enough
to put bounds on the bias evolution from theory, using models and/or simulations to predict the galaxy bias evolution
as a function of redshift, luminosity, color, etc.
In this work however, we will simply quantify what level of knowledge of the galaxy bias evolution
is required for the cross-correlation method to benefit weak lensing studies.
These results can then be used as a target for whatever method (or combination of methods) to
determine the galaxy bias is used.

Finally, for the galaxy bias of the {\it spectroscopic} sample, we again assume a scale independent bias described by a free parameter,
$b^{(s)}_j$ for each spectroscopic galaxy bin.
We will not impose external priors on the spectroscopic galaxy bias because it can be measured quite accurately
using the power spectra of the spectroscopic sample (since there the redshift distribution is assumed to be known perfectly).
We describe our (survey dependent) choice of binning in the next section.

\section{Spectroscopic Redshift Surveys}
\label{sec:spec surveys}

\begin{figure*}
  \begin{center}{
  \includegraphics[width=0.48\columnwidth]{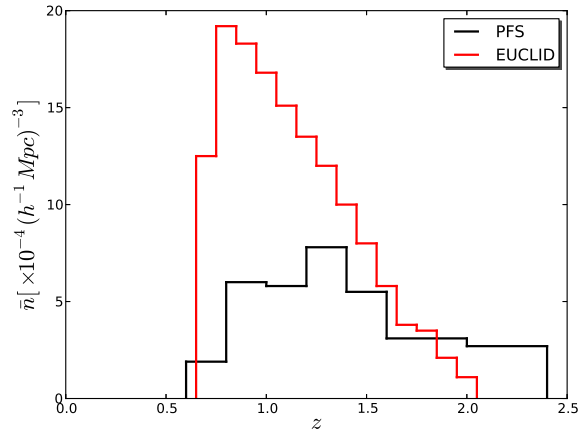}
  }
  \end{center}
  \caption{Assumed comoving number density of spectroscopic galaxies as a function of redshift for
  the Prime Focus Spectrograph (PFS) survey (black) and for EUCLID (red).
}
  \label{fig:specdist}
\end{figure*}

We consider two upcoming galaxy redshift surveys that overlap with the lensing surveys discussed in Section \ref{sec:wlsurvey}.
For HSC, we study the Prime Focus Spectrograph (PFS) cosmology survey (\cite{pfsreport12}), also with the Subaru telescope,
and planned to start in early 2018. Together, these surveys are known as Subaru Measurement of Images and Redshifts (SuMIRe).
EUCLID has its own redshift survey, of which we study the complementarity with its lensing survey.
Our forecasts of the spectroscopic surveys are based on the specifications in \cite{pfsreport12} (PFS) and
\cite{amendolaetal12} (EUCLID),
We assume the same sky coverage for each spectroscopic survey as for its matching lensing survey (see Table \ref{tab:wlspecs}).
Figure \ref{fig:specdist} depicts the assumed comoving number density as a function of redshift
for each survey.

For the fiducial galaxy bias, we follow Table 2 of \cite{pfsreport12} for PFS, and $b^{(s)} = \sqrt{1 + z}$ for EUCLID.
We use the binning $z = 0.6 - 0.8 - 1.0 - 1.2 - 1.4 - 1.6 - 2.0 - 2.4$ (7 bins) for PFS and
$z = 0.65 - 0.75 - 0.85 - 0.95 - 1.05 - 1.15 - 1.25 - 1.35 - 1.45 - 1.55 - 1.65 - 1.75 - 1.85 - 1.95 - 2.05$ (14 bins)
for EUCLID. We remind the reader that each redshift bin has an independent spectroscopic galaxy bias parameter
associated with it. Finally, we need to specify the bins that define the {\it photometric} galaxy bias
parameters. Here we choose $z = 0.0 - 0.6 - 0.8 - 1.0 - 1.2 - 1.4 - 1.6 - 2.0 - 2.4 - 4$
for both survey (i.e.~coinciding
with the PFS spectroscopic bins, but adding a bin at the low and high redshift ends).
Note that the bin widths for the $b^{(p)}(z)$ are typically smaller than the tomographic redshift
bin widths to allow for as general as possible a redshift dependence of $b^{(p)}(z)$.

The redshift bins above are our default choices. We will also consider the effect of varying these choices
and will show that our results are robust with respect to the details of the binning of the spectroscopic sample and
the binning defining the galaxy bias evolution.

When including galaxy clustering, we apply a cutoff to avoid using modes that are too far into
the non-linear regime, $\ell_{{\rm max},i} = k_{\rm max} \cdot D(z_i)$, where $D(z_i)$ is the comoving angular diameter distance
(as in Section \ref{sec:shear th})
to the central redshift of the $i$-th bin. Our standard choice for the comoving wave vector is $k_{\rm max} = 0.2 h/$Mpc,
but we will study in detail the $k_{\rm max}$ dependence of our results. We keep the cutoff in the cosmic shear analysis constant
at $\ell_{\rm max} = 2000$ unless explicitly stated otherwise.

\section{Results of Cross-Correlations Technique}
\label{sec:cc results}

\subsection{SuMIRe}
\label{subsec:sumire}

\subsubsection{photo-$z$ calibration using cross-correlations}

We now use the galaxy clustering information in SuMIRe, i.e.~the $7 \cdot 7 = 49$ $sp$ cross-spectra,
the $\ha 7 (7 + 1) = 28$ $ss$ auto-spectra (although only the 7 auto-spectra actually contain information because of the absence
of overlap between the spectroscopic bins) and the $\ha 7 (7 + 1) = 28$ $pp$ auto-spectra,
to constrain the $22$ photo-$z$ parameters $\{\sigma_{z,i}, b_{z,i}\}$. We marginalize over the cosmological
parameters in the process. Because of the large freedom in the redshift evolution of $\sigma_z(z)$ and $b_z(z)$,
the resulting uncertainties in the individual photo-$z$ parameters are very large ($\sigma(\sigma_{z,i})$ and $\sigma(b_{z,i}) \gg 1$). However, this does not necessarily mean
there is no useful photo-$z$ information in the galaxy clustering spectra. What matters is how well the $ps+pp+ss$ constrain
the linear combinations of photo-$z$ parameters that are degenerate with the effect of cosmological parameters on
the lensing spectra. It is possible for these parameter directions to be well (enough) constrained while the individual
$\sigma_{z,i}$ and $b_{z,i}$ parameters have large error bars.

We next consider explicitly to what extent the photo-$z$ information from $ps+pp+ss$ helps the weak lensing (+Planck) analysis
of cosmological parameters. In Table \ref{tab:results SuMIRe}, we repeat in the first and last column the cases discussed
in Section \ref{sec:results wl} of free photo-$z$ parameters (no external priors) and exactly known photo-$z$ parameters, respectively.
These are the two extreme cases to which we can compare the results using the cross-correlation technique. Ideally,
adding the photo-$z$ information from $ps + pp + ss$ will bring the uncertainties and FOM close to the case of perfectly known
photo-$z$ parameters. What we actually find is shown in the two central columns. The second column shows the results when the galaxy bias
of the photometric (source) sample is known perfectly. There is clear improvement, with the dark energy FOM increasing by more than a factor
four. However, the result is a long way off from the case of perfectly known photo-$z$ parameters.
In the third column, we consider the constraints when the redshift evolution of the photometric galaxy bias is
unknown {\it a priori}
(i.e.~self-calibrated by the $ps+pp+ss$ data). We see that leaving $b^{(p)}(z)$ free deteriorates constraints,
but only by about $25 \%$.

\begin{table*}[hbt!]
\begin{center}
\small
\begin{tabular}{c|cccc}
\hline\hline
$\sigma(p)$ & $\gamma \gamma$ (``free'' $\sigma_z, b_z$) & $\gamma \gamma$ + $ps + pp + ss$ ($b^{(p)}(z)$ known) & $\gamma \gamma$ + $ps + pp + ss$ ($b^{(p)}(z)$ unknown) &  $\gamma \gamma$ (known $\sigma_z, b_z$)  \\
\hline\hline
$\omega_b$ & 0.00013 & 0.00012 & 0.00012 & 0.00011 \\
$\omega_c$ & 0.0011 & 0.0010 & 0.0011 & 0.00069 \\
$\Omega_\Lambda$ & 0.17 & 0.11 & 0.12 & 0.051 \\
$n_s$ & 0.0033 & 0.0031 & 0.0032 & 0.0027 \\
$\sigma_8$ & 0.18 & 0.11 & 0.12 & 0.043 \\
$w_0$ & 1.5 & 1.0 & 1.2 & 0.61 \\
$w_a$ & 3.7 & 2.4 & 2.7 & 1.6 \\
FOM$=1/\sqrt{{\rm Det Cov}}$ & 0.52 & 2.3 & 1.7 & 9.5 \\
\hline\hline
\end{tabular}
\caption{Forecasted constraints for SuMIRe (HSC lensing + PFS galaxy clustering),
using the cross-correlation method to calibrate photo-$z$ parameters.
The far left and far right columns show the extreme cases where the galaxy clustering information
($ps+pp+ss$) is not used and the photo-$z$ parameters are either assumed unknown {\it a priori} (far left),
or known exactly (far right). The two columns in the middle include the $ps+pp+ss$ information
and assume no prior knowledge on the photo-$z$ parameters. The cross-correlation method thus helps
improve relative to the case of {\it a priori} unknown photo-$z$ parameters, but is not as good as the case
where there is no uncertainty in the shape of the photo-$z$ distribution.
All results shown include a Planck prior.
}
\label{tab:results SuMIRe}
\end{center}
\end{table*}

\subsubsection{including information from direct photo-$z$ calibration}

It is instructive to compare the yield of the cross correlation technique to imposing a simple diagonal prior on
the photo-$z$ parameters.
This prior represents the level of calibration that has been achieved for the photo-$z$ estimator.
For simplicity, we consider the case where we can place a constant prior on all photo-$z$
parameters, $\sigma_{\rm prior}(\sigma_{z,i}) = \sigma_{\rm prior}(b_{z,i}) \equiv \sigma_{\rm prior}$ for all $i$.
We find that the obtained figure of merit in Table \ref{tab:results SuMIRe} of FOM $= 2.3 (1.7)$ for known (unknown) photometric galaxy bias
can also be reached with a prior $\sigma_{\rm prior} = 0.04 (0.05)$. It is not unlikely that photometric redshifts will
be calibrated to this level so that using the $ps+pp+ss$ galaxy clustering information in the absence of a prior
on the photo-$z$ parameters is not better than having a prior and not using the information from $ps+pp+ss$ at all.
However, this is not the proper comparison to make. In reality, there will always be some prior on the photo-$z$ parameters
and we should ask the question how much improvement one gets from adding $ps+pp+ss$.

We address this question in
Table \ref{tab:ext prior sumire}, where we show the dark energy figure of merit with and without $ps+pp+ss$ for
different priors on the photo-$z$ parameters. We find that if $\sigma_{\rm prior}$ is larger than $\sim 0.01$, adding
galaxy clustering information causes a significant improvement in FOM, but if the prior is better than this,
the cross-correlation technique does not add much.

When $ps+pp+ss$ improves constraints, there is a significant advantage to
knowing the galaxy bias. For instance, if the photo-$z$ prior is $\sigma_{\rm prior} = 0.05$,
the FOM is 7.0 if $b^{(p)}(z)$ is assumed to be known and 4.7 when it is left free.
We tested how well the galaxy bias needs to be known {\it a priori} to improve from FOM$=4.7$
to (close to) FOM $ = 7.0$. Applying an independent prior to each bin (see the first bullet point in the
discussion in the end of Section \ref{sec:cc th} for the definition of the prior),
we find that a percent level prior $\sigma_{\rm diag}^{\rm bias}$
significantly improves the dark energy FOM. For example, $\sigma_{\rm diag}^{\rm bias} = 0.02$
gives FOM$=5.7$ (approximately halfway between the cases of unknown and known galaxy bias),
and $\sigma_{\rm diag}^{\rm bias} = 0.01$
gives FOM$=6.3$.
Using our second type of galaxy bias prior (the second bullet point in the end of Section \ref{sec:cc th}),
we find that any prior $\sigma_{\rm lin}^{\rm bias}$ brings the figure of merit virtually all the way to its optimal value
FOM$=7.0$. In other words, merely imposing that the galaxy bias is linear in redshift constrains the bias evolution
sufficiently for it to not hinder the determination of the photo-$z$ parameters using $ps+pp+ss$.

\begin{table*}[hbt!]
\begin{center}
\small
\begin{tabular}{c|ccc}
\hline\hline
prior on $\sigma_{z,i}, b_{z,i}$ & $\gamma \gamma$ & $\gamma \gamma + ps + pp + ss$ (known $b^{(p)}(z)$) & $\gamma \gamma + ps + pp + ss$ (unknown $b^{(p)}(z)$) \\
\hline\hline
no prior & 0.52 & 2.3 & 1.7 \\
0.05 & 1.8 & 7.0 & 4.7 \\
0.02 & 4.4 & 7.5 & 6.0 \\
0.01 & 7.0 & 8.2 & 7.6 \\
0.005 & 8.7 & 8.9 & 8.8 \\
0.0 & 9.5 & 9.5 & 9.5 \\
\hline\hline
\end{tabular}
\caption{Dark energy figure of merit for SuMIRe
as a function of the prior knowledge of the photo-$z$ parameters.
Columns show results for shear only (left) and shear with $ps+pp+ss$ (middle and right).
Depending on the prior on the photo-$z$ parameters, the cross-correlation method
can help strongly improve the dark energy constraint relative to the case with shear information only.
All results shown include a Planck prior.
}
\label{tab:ext prior sumire}
\end{center}
\end{table*}

\subsubsection{dependence on $k_{\rm max}$ and on modeling of photo-$z$ distribution}

\begin{figure*}
  \begin{center}{
  \includegraphics[width=0.48\columnwidth]{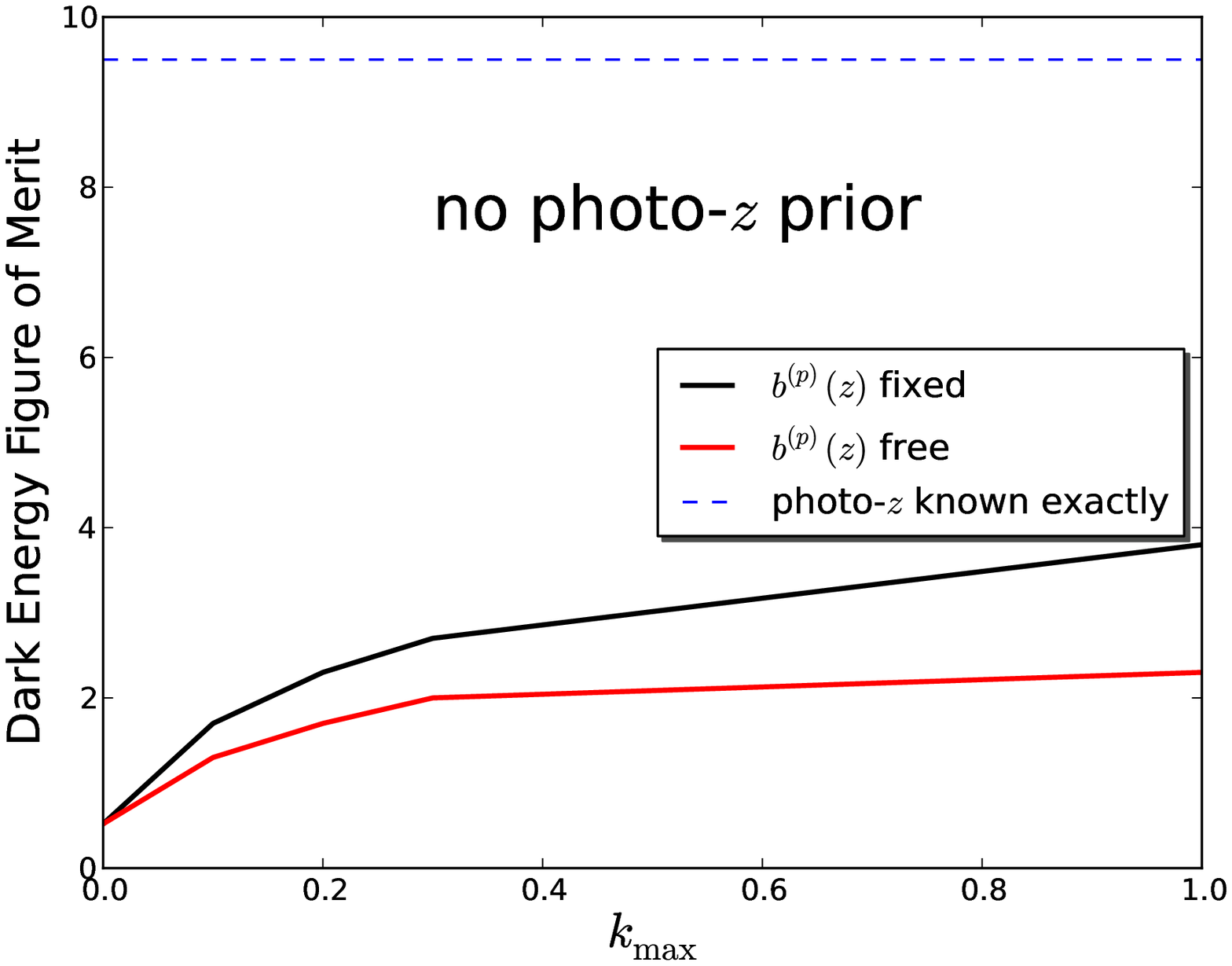}
  \includegraphics[width=0.48\columnwidth]{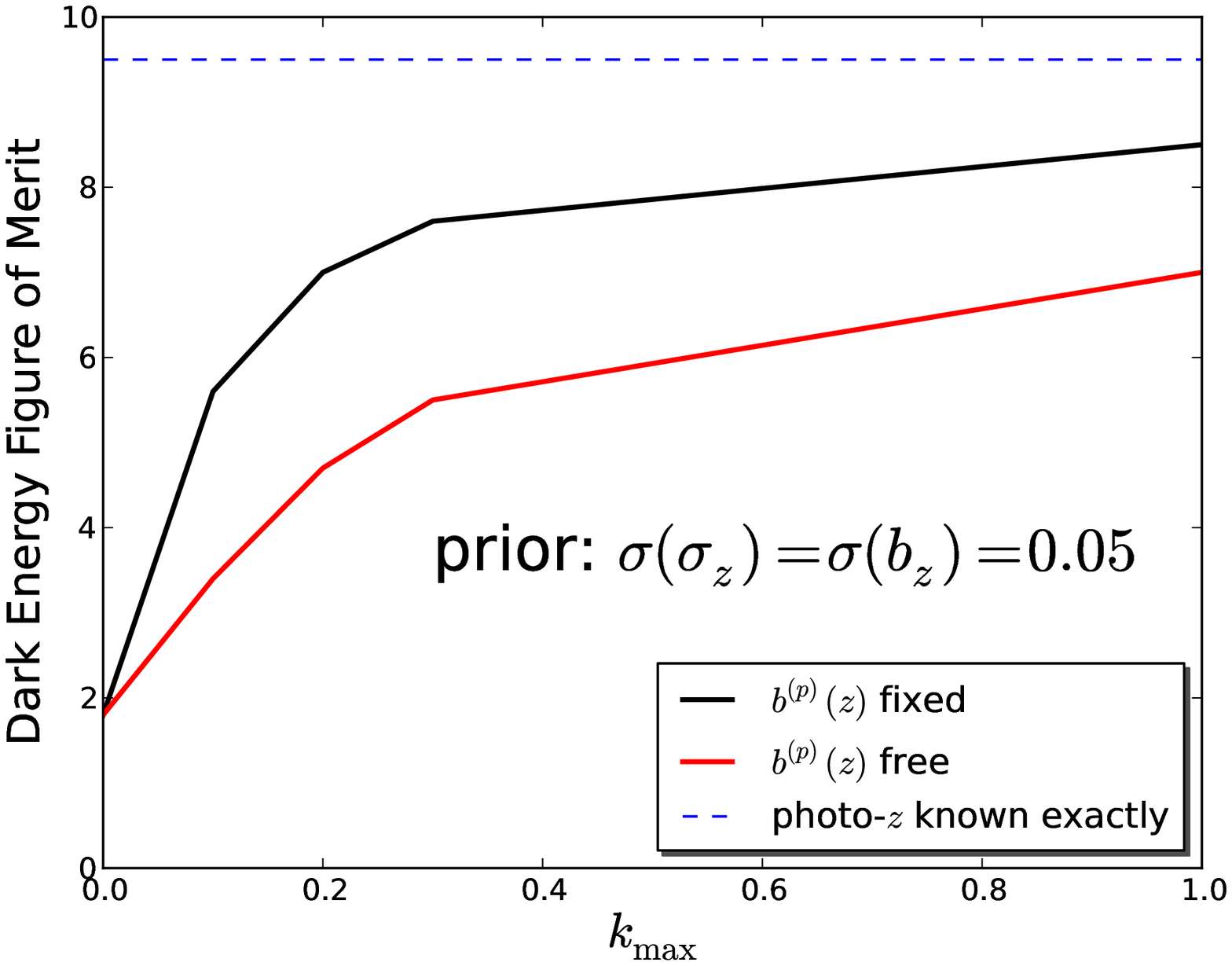}
  \includegraphics[width=0.48\columnwidth]{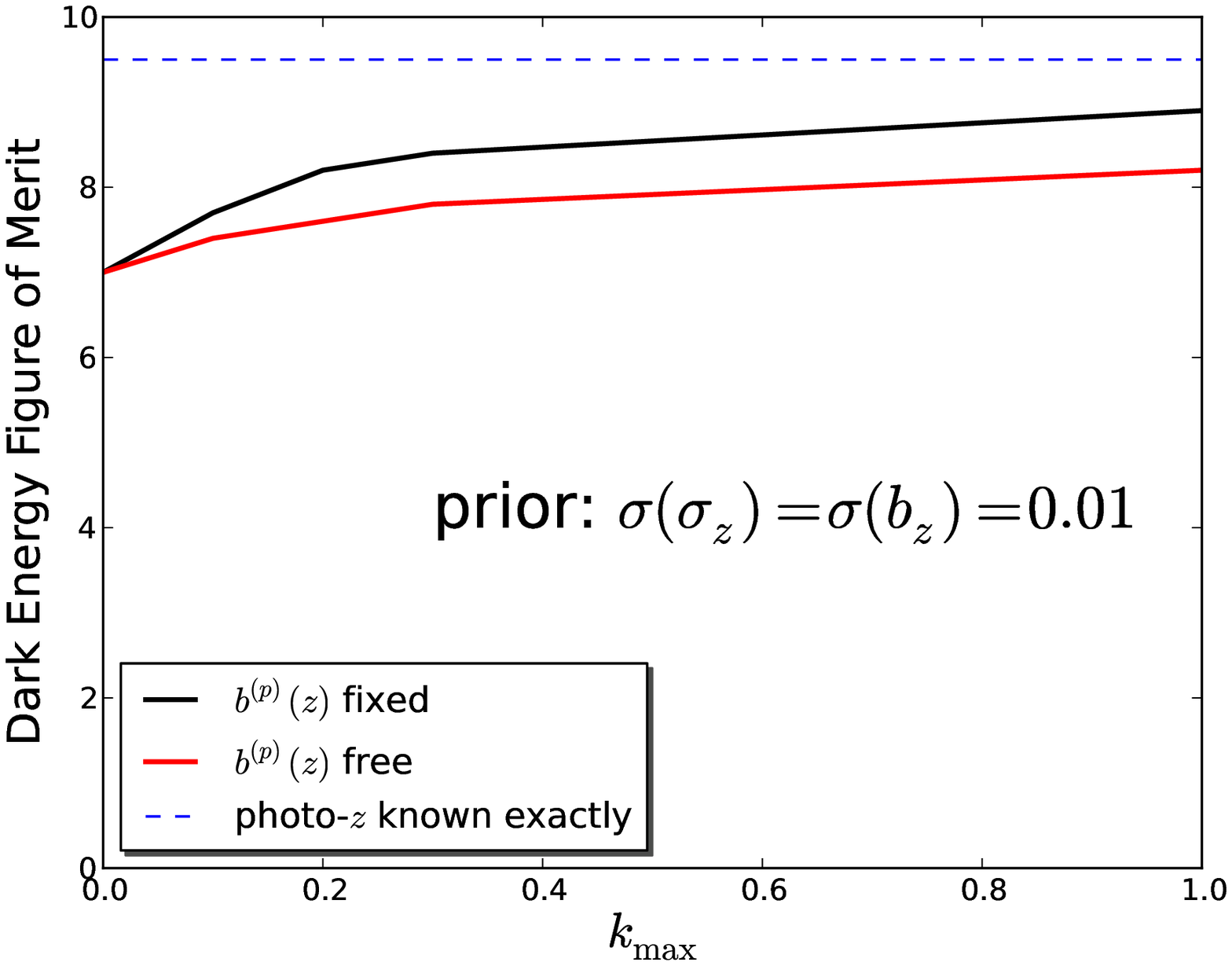}
  }
  \end{center}
  \caption{Dependence of the SuMIRe dark energy figure of merit on
  the largest wave vector, $k_{\rm max}$, included in the analysis
  of the cross- and auto-spectra ($ps+pp+ss$). Our default choice is $k_{\rm max} = 0.2 h/$Mpc.
  Top left: no photo-$z$ prior, top right: $\sigma_{\rm prior} = 0.05$,
  bottom: $\sigma_{\rm prior} = 0.01$.
  Results are shown for the case of known photometric galaxy bias
  and of free (self-calibrated) galaxy bias.
  In all cases, the dependence of the photo-$z$ information
  (which is the information from $ps+pp+ss$ that we focus on in this work and that drives the improvement
  in weak lensing constraints on dark energy)
  on $k_{\rm max}$ past
  $k_{\rm max} = 0.2 h/$Mpc is rather weak, showing that there is not much to gain from
  pushing the analysis to smaller scales.
  All results shown include a Planck prior.
  }
  \label{fig:fom vs kmax sumire}
\end{figure*}

We now consider the dependence of the dark energy figure of merit on
the maximum wave vector, $k_{\rm max}$, that is included in the $ps+ss+pp$.
We keep the range of scales used for the lensing analysis fixed ($\ell_{\rm max} = 2000$).
The results (again for both known galaxy bias and free galaxy bias) are shown in Figure \ref{fig:fom vs kmax sumire}.
The top left panel gives the figure of merit in the absence of any prior on the photo-$z$ parameters,
the top right panel describes the case of a (very modest) prior $\sigma_{\rm prior} = 0.05$,
and the bottom plot is for a stronger prior $\sigma_{\rm prior} = 0.01$. In each case, the $k_{\rm max}$
dependence beyond our default choice $k_{\rm max} = 0.2 h/$Mpc is not very strong.

Finally, we wish to point out that the success of the cross-correlation method depends strongly on
how much freedom is allowed in the photo-$z$ distribution and its evolution. In the above,
we have chosen a fairly general approach with a total of 22 photo-$z$ parameters. We now briefly consider
the results when
we assume the photo-$z$ distribution $p(z_{\rm ph}|z)$ to be redshift independent, i.e.~the parameters
$\sigma_z$ and $b_z$ are constants. We choose a fiducial $\sigma_z =0.1$ (equal to the value of $\sigma(z)$
in the previous redshift dependent model at the average redshift $z=1$) and
$b_z = 0$, and now have only two free photo-$z$ parameters. In this case,
the shear-only FOM$=1.9$ is already almost four times as large as in the case of redshift
dependent photo-$z$ parameters (fixing the photo-$z$ parameters gives FOM$=9.3$, which is similar to the original result,
as expected). Adding the information from $ps+pp+ss$, lifts the figure of merit to FOM$=8.9 (8.6)$ for fixed (free)
photometric galaxy bias. In this more restricted photo-$z$ model, the cross-correlation method is thus significantly more
powerful. Moreover, the constraints from $ps+pp+ss$ on the individual photo-$z$ parameters
are now quite strong (unlike in the more general model), $\sigma(b_z) = 0.004$ and $\sigma(\sigma_z) = 0.004$ (here,
no galaxy bias prior is assumed).

\subsection{EUCLID}
\label{subsec:euclid}

\begin{table*}[hbt!]
\begin{center}
\small
\begin{tabular}{c|cccc}
\hline\hline
$\sigma(p)$ & $\gamma \gamma$ (``free'' $\sigma_z, b_z$) & $\gamma \gamma$ + $ps + pp + ss$ ($b^{(p)}(z)$ known) & $\gamma \gamma$ + $ps + pp + ss$ ($b^{(p)}(z)$ unknown) &  $\gamma \gamma$ (known $\sigma_z, b_z$)  \\
\hline\hline
$\omega_b$ & 0.00013 & 0.000096 & 0.00011 & 0.000093 \\
$\omega_c$ & 0.0011 & 0.00040 & 0.00065 & 0.00031 \\
$\Omega_\Lambda$ & 0.13 & 0.025 & 0.050 & 0.013 \\
$n_s$ & 0.0031 & 0.0022 & 0.0025 & 0.0021 \\
$\sigma_8$ & 0.12 & 0.021 & 0.042 & 0.011 \\
$w_0$ & 1.4 & 0.27 & 0.54 & 0.14 \\
$w_a$ & 3.6 & 0.67 & 1.3 & 0.36 \\
FOM$=1/\sqrt{{\rm Det Cov}}$ & 1.5 & 62 & 26 & 162 \\
\hline\hline
\end{tabular}
\caption{Forecasted constraints for EUCLID,
using the cross-correlation method to calibrate photo-$z$ parameters.
The far left and far right columns show the extreme cases where the galaxy clustering information
($ps+pp+ss$) is not used and the photo-$z$ parameters are either assumed unknown {\it a priori} (far left),
or known exactly (far right). The two columns in the middle include the $ps+pp+ss$ information
and assume no prior knowledge on the photo-$z$ parameters. The cross-correlation method thus helps
improve relative to the case of {\it a priori} unknown photo-$z$ parameters, but is not as good as the case
where there is no uncertainty in the shape of the photo-$z$ distribution.
All results shown include a Planck prior.
}
\label{tab:results EUCLID}
\end{center}
\end{table*}

\subsubsection{photo-$z$ calibration using cross-correlations}

We now consider the photo-$z$ information in the EUCLID lensing source sample and the EUCLID spectroscopic
galaxy sample. Using the cross- and auto-spectra $ps+pp+ss$, we find strong direct constraints on
a large number of the $\sigma_{z,i}$ and $b_{z,i}$ parameters, with the uncertainties in the best measured nodes
$\sigma(b_z) \sim 0.001$ and $\sigma(\sigma_z) \sim 0.002$ (assuming no galaxy bias prior).
Table \ref{tab:results EUCLID} (cf.~Table \ref{tab:results SuMIRe}) shows the effect of the photo-$z$ prior from $ps+pp+ss$ on the cosmological
constraints from cosmic shear (+Planck).
We again find that the cross-correlation technique significantly improves the weak lensing bounds
relative to the case of ({\it a priori}) unknown photo-$z$ parameters, with the dark energy
figure of merit increasing from $1.5$ to $62 (26)$ for known (unknown) source galaxy bias.
As was the case for SuMIRe, the constraints with $ps+pp+ss$ are still not at the level of cosmic shear
with perfectly known source distribution. Moreover, not knowing the bias $b^{(p)}(z)$ aversely affects the constraints
(decreasing the FOM by $\sim 60 \%$).

\begin{table*}[hbt!]
\begin{center}
\small
\begin{tabular}{c|ccc}
\hline\hline
prior on $\sigma_{z,i}, b_{z,i}$ & $\gamma \gamma$ & $\gamma \gamma + ps + pp + ss$ (known $b^{(p)}(z)$) & $\gamma \gamma + ps + pp + ss$ (unknown $b^{(p)}(z)$) \\
\hline\hline
no prior & 1.5 & 62 & 26 \\
0.05 & 3.9 & 101 & 61  \\
0.02 & 10 & 106 & 75  \\
0.01 & 26 & 114 & 96 \\
0.005 & 61 & 127 & 119 \\
0.002 & 124 & 145 & 143 \\
0.001 & 150 & 154 & 153  \\
0.0 & 162 & 162 & 162 \\
\hline\hline
\end{tabular}
\caption{Dark energy figure of merit for EUCLID
as a function of the prior knowledge of the photo-$z$ parameters.
Columns show results for shear only (left) and shear with $ps+pp+ss$ (middle and right).
Depending on the prior on the photo-$z$ parameters, the cross-correlation method
can help strongly improve the dark energy constraint relative to the case with shear information only.
All results shown include a Planck prior.
}
\label{tab:ext prior euclid}
\end{center}
\end{table*}

\subsubsection{including information from direct photo-$z$ calibration}

The gains from the cross-correlation method shown in Table \ref{tab:results EUCLID}, i.e.~FOM$=62 (26)$,
are equivalent to having a photo-$z$ prior $\sigma_{\rm prior} = 0.005 (0.010)$, showing the strength of this technique
for a survey like EUCLID.
In Table \ref{tab:ext prior euclid}, we show the constraints with and without use of $ps+pp+ss$
for various external priors on the photo-$z$ parameters.
Unless this prior is very strong $\sigma_{\rm prior} < 0.002$, the cross-correlation method always
helps to strongly improve the cosmological constraints.

Considering as an example the case of a photo-$z$ prior $\sigma_{\rm prior} = 0.05$ (as we did for SuMIRe)
the figure of merit ranges from FOM $= 61 - 101$, depending on how much information on $b^{(p)}(z)$
is available. We find that a diagonal bias prior $\sigma^{\rm bias}_{\rm diag} = 0.005$ gives a FOM
halfway between these two extremes (FOM$=81$), while an even stronger prior $\sigma^{\rm bias}_{\rm diag} \leq 0.002$
is essentially equivalent to knowing the galaxy bias perfectly,
lifting the figure of merit to FOM$\geq 94$. The bias knowledge requirements are thus stricter than in the case of SuMIRe.
On the other hand, imposing $b^{(p)}(z)$ to be linear in $z$ is already enough to ensure FOM$\geq 100$
even if no prior is imposed on the slope of this relation (i.e.~$\sigma^{\rm bias}_{\rm lin}$).

\begin{figure*}
  \begin{center}{
  \includegraphics[width=0.48\columnwidth]{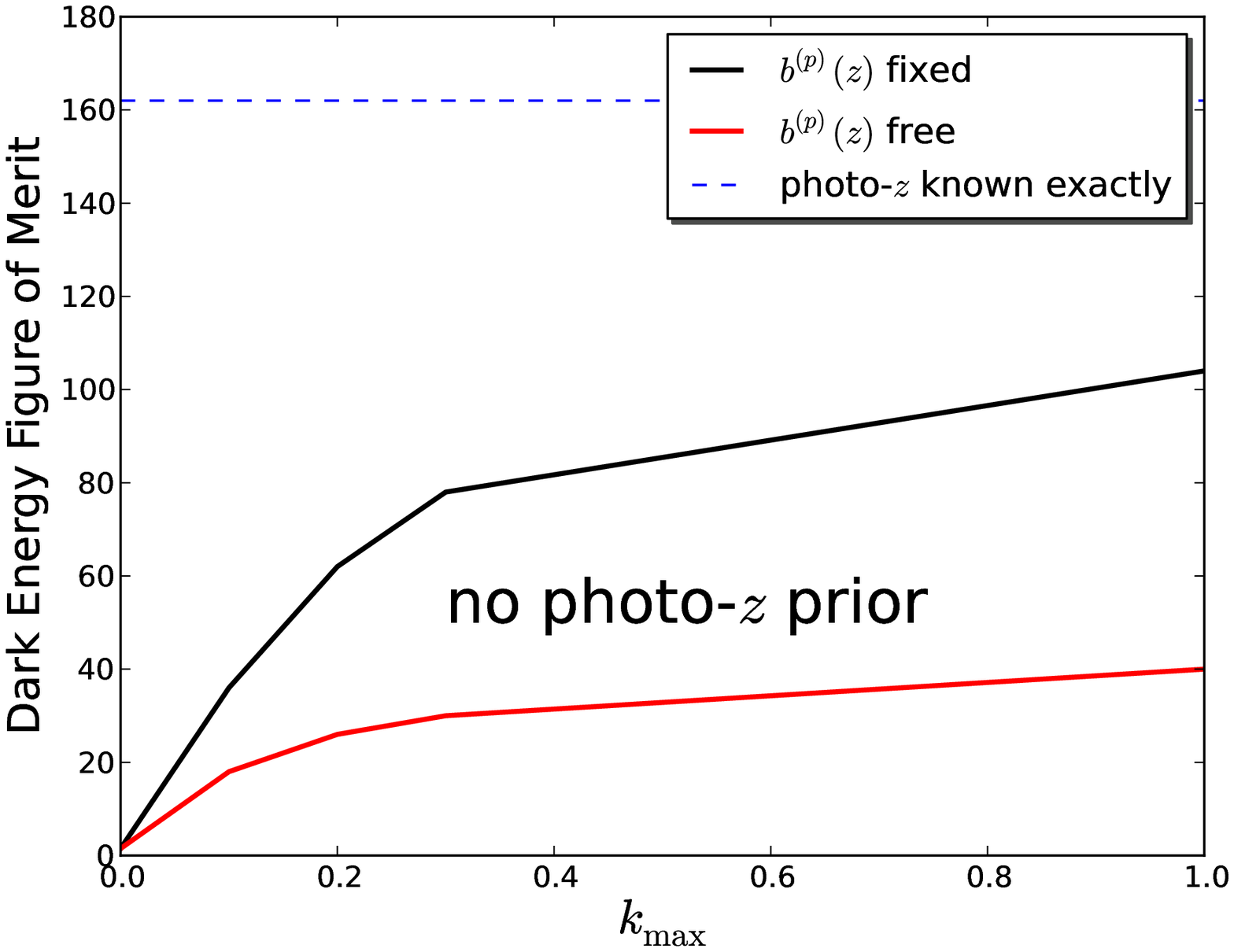}
  \includegraphics[width=0.48\columnwidth]{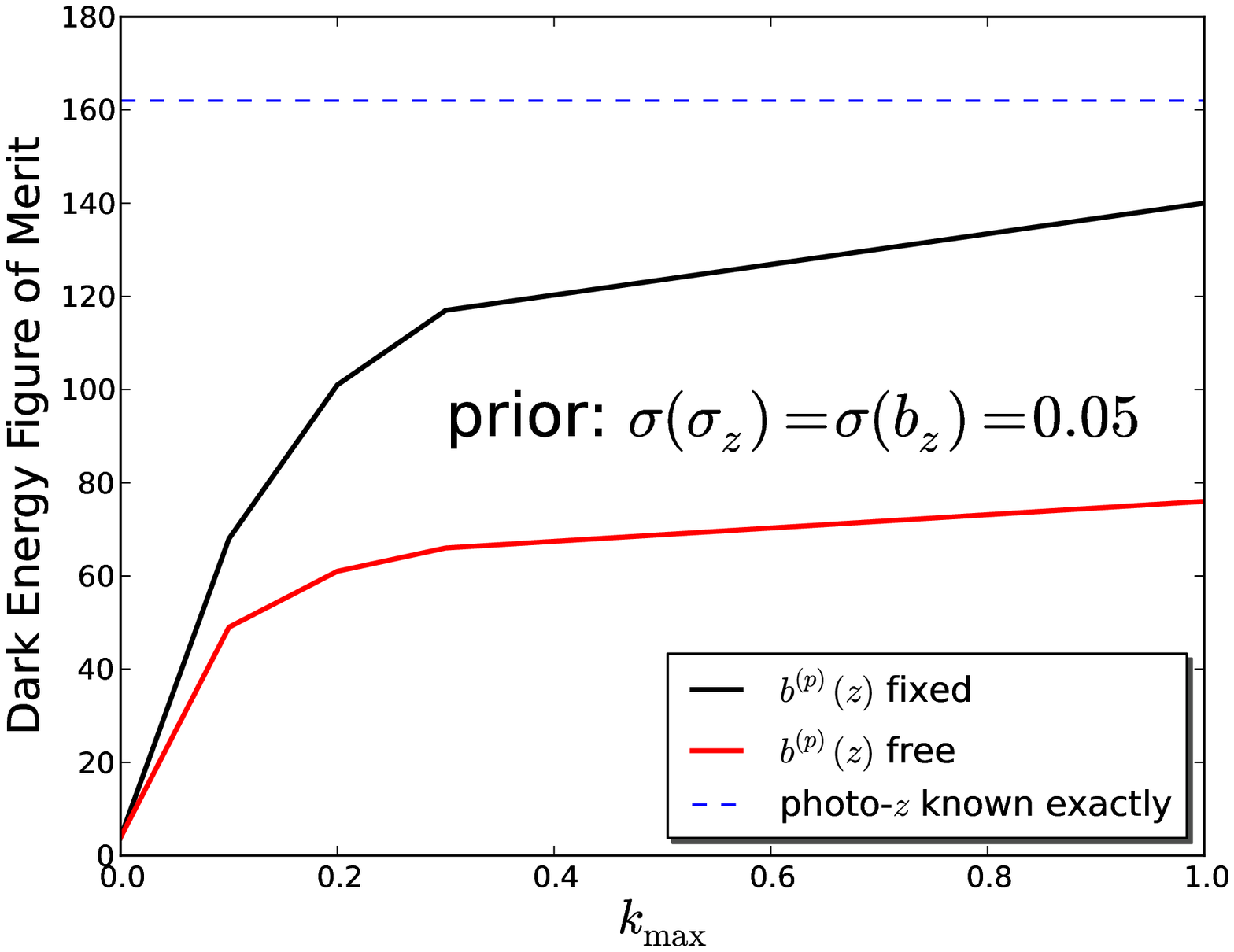}
  \includegraphics[width=0.48\columnwidth]{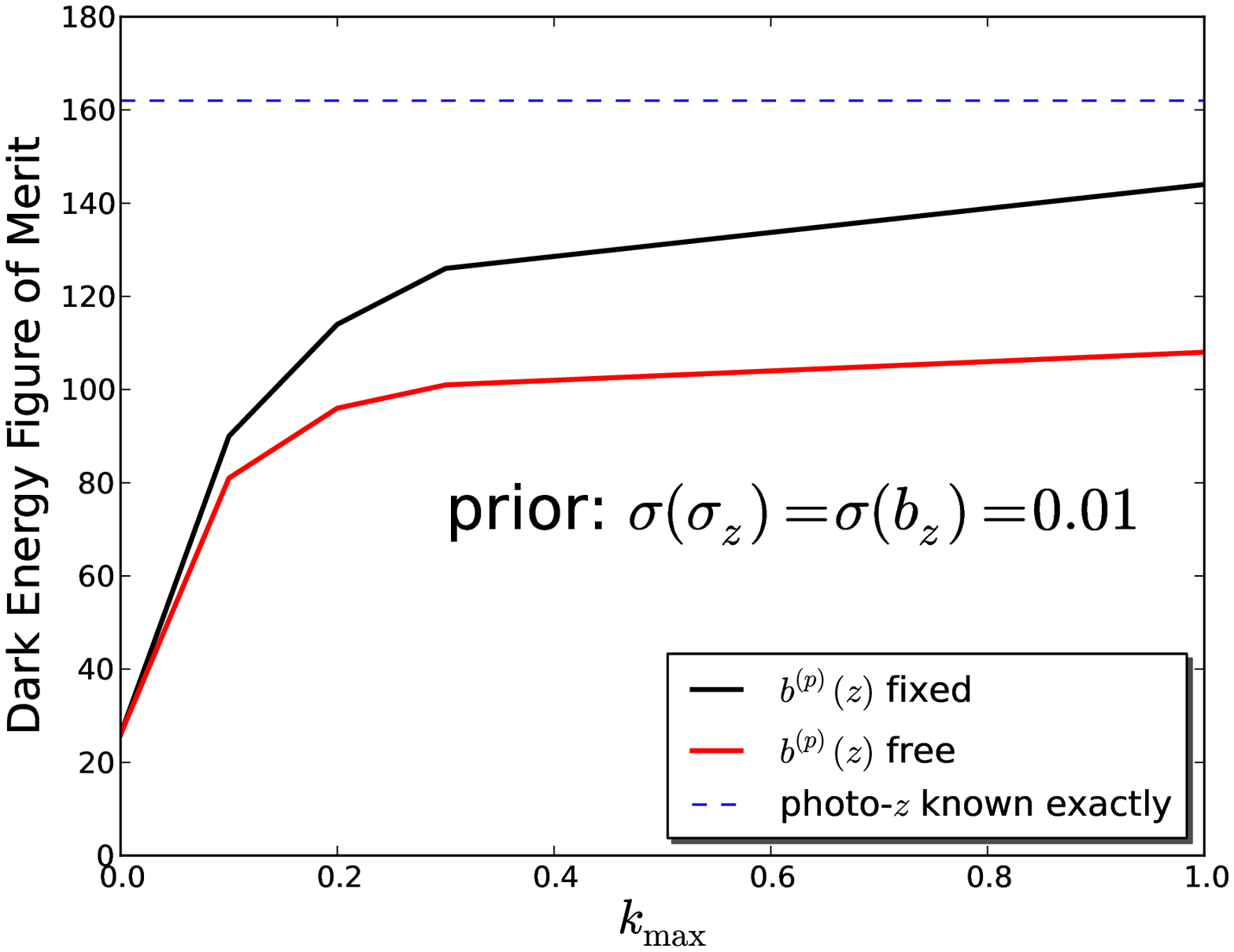}
  }
  \end{center}
  \caption{Dependence of the EUCLID dark energy figure of merit on
  the largest wave vector, $k_{\rm max}$, included in the analysis
  of the cross- and auto-spectra ($ps+pp+ss$). Our default choice is $k_{\rm max} = 0.2 h/$Mpc.
  Top left: no photo-$z$ prior, top right: $\sigma_{\rm prior} = 0.05$,
  bottom: $\sigma_{\rm prior} = 0.01$.
  Results are shown for the case of known photometric galaxy bias
  and of free (self-calibrated) galaxy bias.
  In all cases, the dependence of the photo-$z$ information
  (which is the information from $ps+pp+ss$ that we focus on in this work and that drives the improvement
  in weak lensing constraints on dark energy)
  on $k_{\rm max}$ past
  $k_{\rm max} = 0.2 h/$Mpc is rather weak, showing that there is not much to gain from
  pushing the analysis to smaller scales.
  All results shown include a Planck prior.
  }
  \label{fig:fom vs kmax euclid}
\end{figure*}

\subsubsection{dependence on $k_{\rm max}$}

The dependence of our forecasts on the smallest modes included in the galaxy clustering analysis is shown in Figure
\ref{fig:fom vs kmax euclid} for several choices of the external photo-$z$ prior $\sigma_{\rm prior}$. As before the $k_{\rm max}$
is not particularly strong beyond our default choice of $k_{\rm max} = 0.2 h/$Mpc.

\section{General redshift distribution (no photo-$z$ information)}
\label{direct}

In the previous sections, we have assumed that photometric redshifts are available for the lensing source
galaxies, taking into account that the photo-$z$ distributions may not be perfectly known.
This uncertainty in the photo-$z$ distribution translates into uncertainty in the lensing source
redshift distribution, which in turn translates into additional uncertainty in cosmological parameters
obtained from cosmic shear tomography (or into parameter bias if the effect is not properly modeled).
The photo-$z$ parameters cannot be properly calibrated by the cosmic shear data itself
and we have shown that the resulting degradation in cosmological parameter uncertainties can be very large,
depending on the prior knowledge of the photo-$z$ parameters.

The main goal of the previous sections (and of this article) was to quantify to what extent cross-correlations between
the source galaxies and an overlapping sample of spectroscopic galaxies (in addition to the autocorrelations
of these samples) can calibrate the photo-$z$ distribution and mitigate the degradation
of cosmological parameter estimation from cosmic shear.
This addresses directly the question of how useful the cross-correlation technique will be
for supporting the constraining power of upcoming lensing surveys.
However, because of the assumption of having photo-$z$'s, and because of the joint analysis with a focus
on the end product (i.e.~cosmological parameters), the analysis thus far does not give much insight into
how well {\it in general} the cross-correlation method can constrain redshift distributions
or what knowledge of the source distribution is required for a lensing analysis. We will
address these questions individually in this section. First, we will consider an {\it a priori}
completely unknown, arbitrary galaxy redshift distribution and study how well it can be measured (Section \ref{subsec:ps general})
using the cross- and autocorrelations of the overlapping galaxy number densities (the $ps+pp+ss$ spectra).
We will not assume any photometric redshift information in this section and will pay specific attention
to the dependence of our results on our knowledge of galaxy bias evolution.
Compared to the rest of this work, Section \ref{subsec:ps general} follows more closely the spirit of previous studies
of the subject of using cross-correlations to constrain (source) redshift distributions, see
\cite{hoetal08,newman08,mathnewman10,schulz10,mcwhite13,menardetal13}.

In section \ref{subsec:lensing req}, we then ask which specific properties (/modes) of the source distribution do we really need to know
in order to not weaken cosmological parameter constraints? Comparing how well $ps+pp+ss$
measures the source distribution to what is the requirement from lensing will then give more insight into
when the cross-correlation technique is useful.

Throughout this section, unless otherwise specified, we assume the survey properties of SuMIRe, as described in
sections \ref{sec:wlsurvey} and \ref{sec:spec surveys}. While the galaxy sample of which we try to measure the redshift distribution
is no longer assumed to have photometric redshift estimates, we will still refer to it as the $p$ sample (and $s$ refers to the spectroscopic sample).
Since for our discussion in Section \ref{subsec:ps general}, this sample also no longer has to be a lensing source sample, we will
no longer refer to it as the photometric or source sample, but instead call it the {\it target} sample.

\subsection{Measuring a general redshift distribution using the cross-correlation technique}
\label{subsec:ps general}

\begin{figure*}
  \begin{center}{
  \includegraphics[width=0.48\columnwidth]{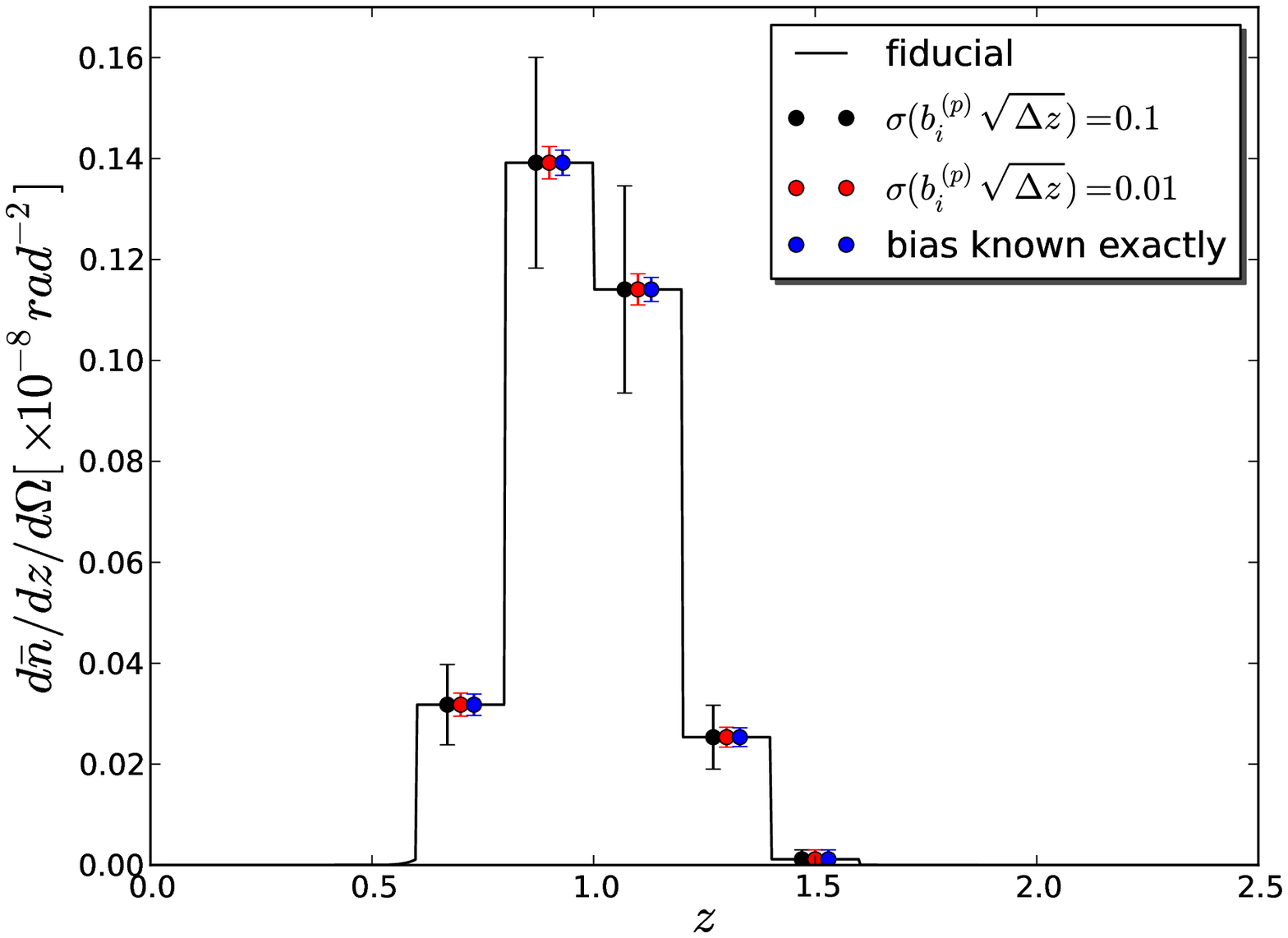}
  \includegraphics[width=0.48\columnwidth]{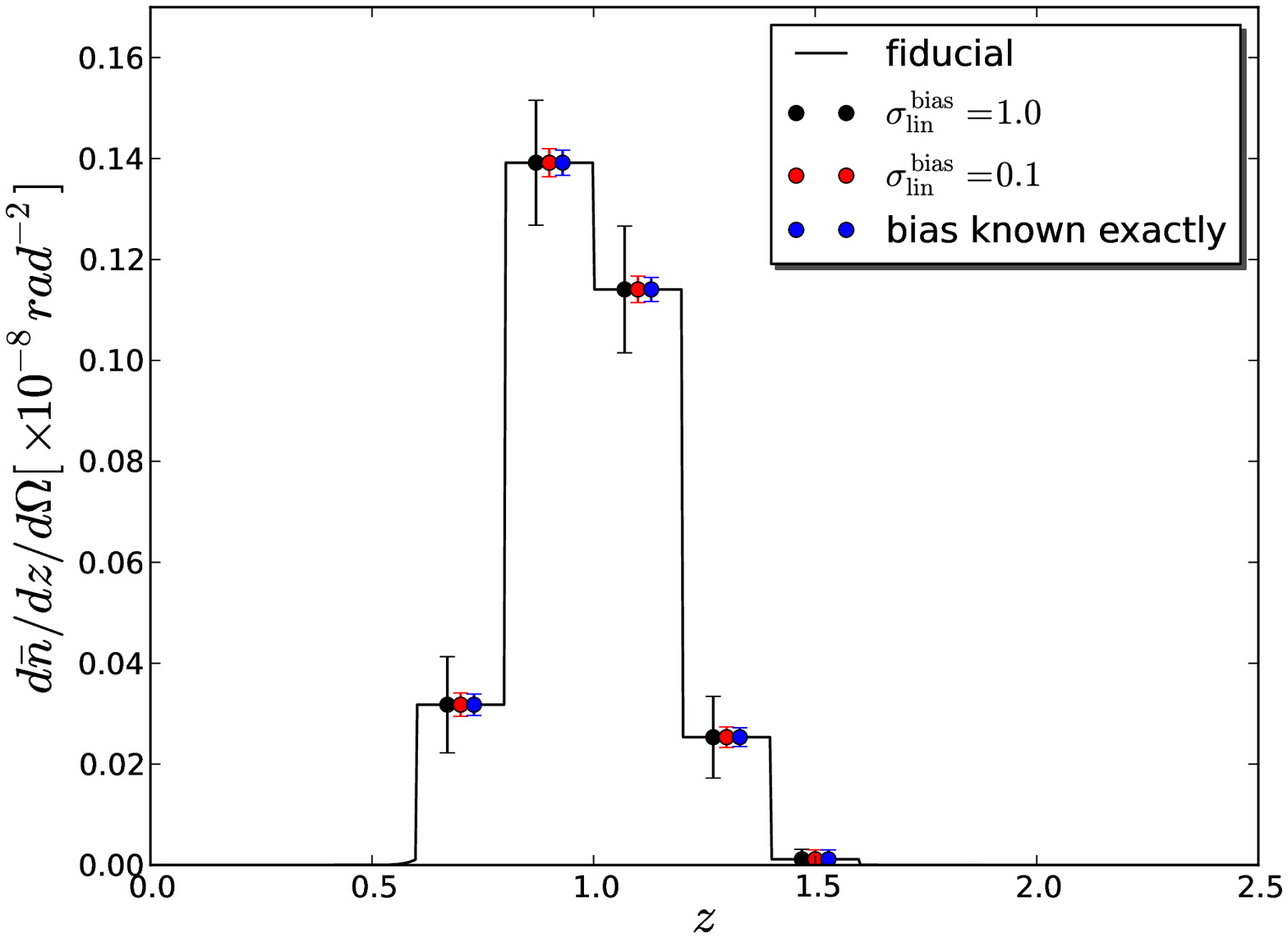}
  \includegraphics[width=0.48\columnwidth]{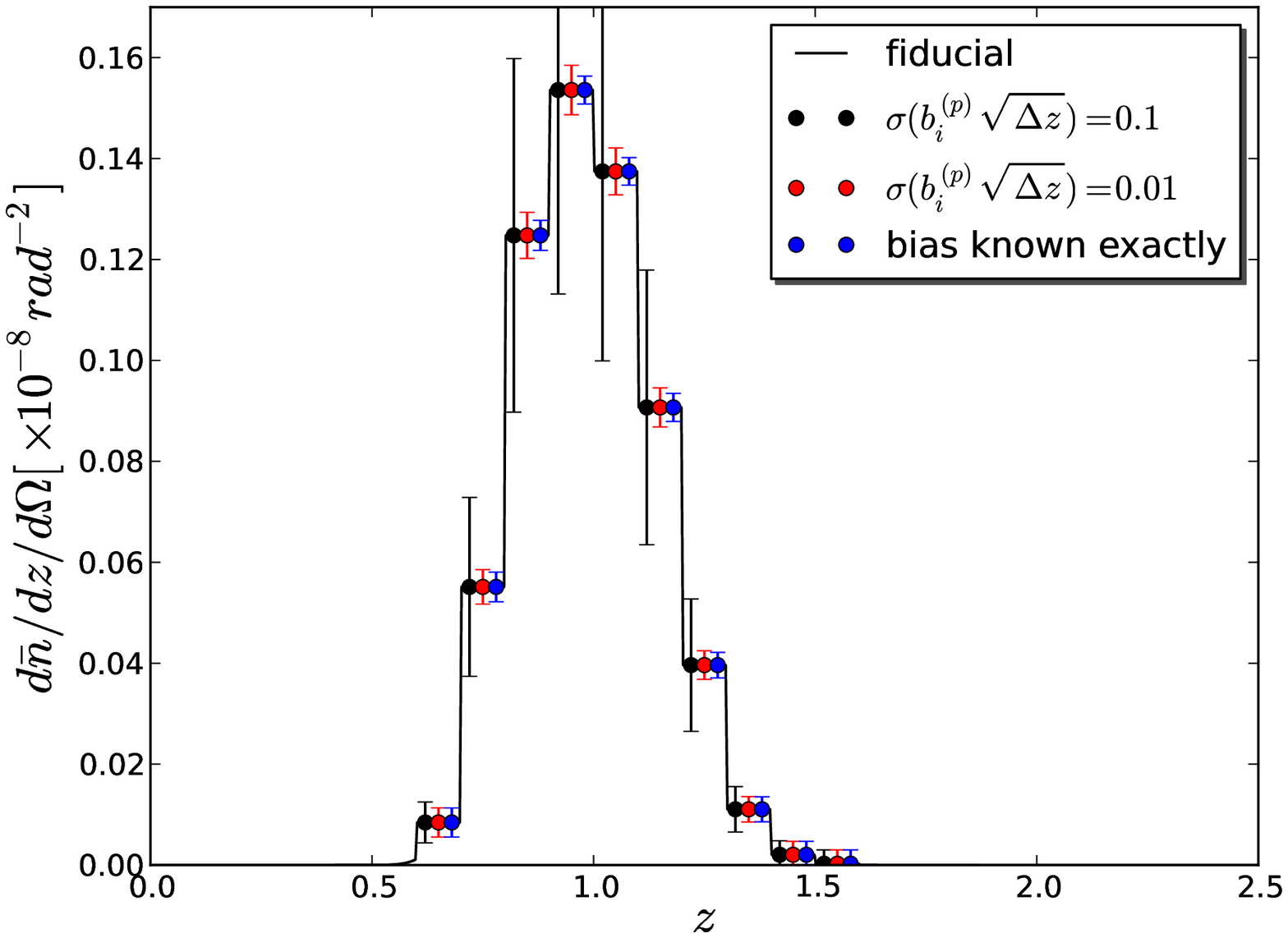}
  \includegraphics[width=0.48\columnwidth]{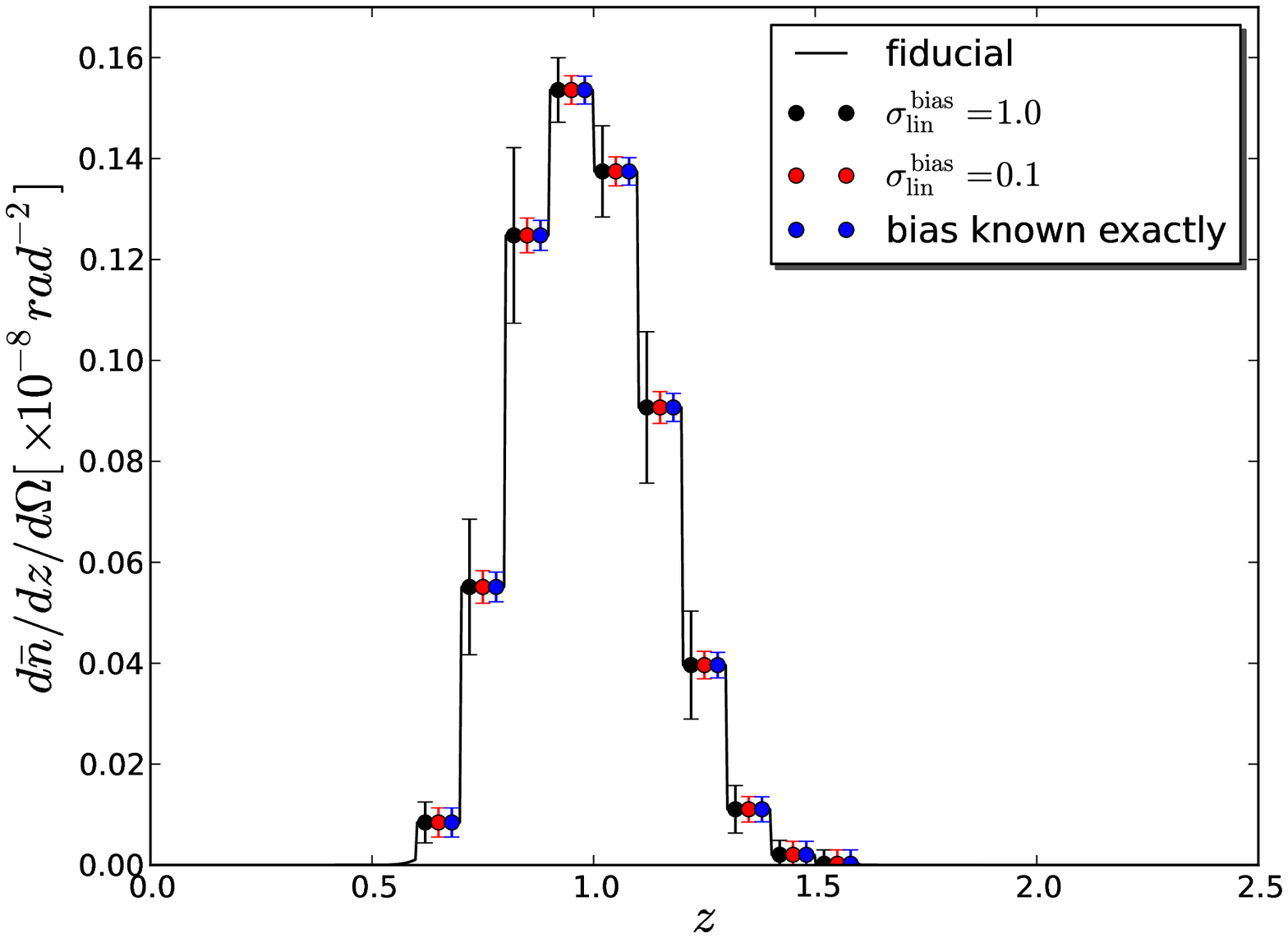}
  }
  \end{center}
  \caption{
  Uncertainties in the redshift distribution after reconstruction with the cross-correlation technique
  (i.e.~using cross-spectra $ps$ and auto spectra $pp$ and $ss$). We assume a HSC-like
  survey (but do not use photometric redshift information). In the top panels,
  the redshift distribution of the $p$ sample is allowed to vary in $N_{dn/dz} = 5$
  redshift bins, while the bottom panels depict the case of $N_{dn/dz}$.
  The results strongly depend on how much knowledge of the galaxy bias $b^{(p)}(z)$
  is assumed. In the left panels, we consider the effect of a diagonal prior on the bias bins, while in the
  right panels, we study the case of a prior on the slope of the bias-redshift relation (see text for details).
  }
  \label{fig:direct dndz}
\end{figure*}

We consider a target sample of galaxies with a distribution based on what would be obtained when
applying a photometric redshift cut
$z_{\rm ph} = 0.8 - 1.2$ to the HSC source sample, with $\sigma_z = 0.05 (1 + z), \, b_z \equiv 0$.
Instead of using the exact resulting distribution, we approximate it by a piecewise constant
function defined in $N_{dn/dz}$ bins. The fiducial distributions are depicted by the black lines
in Fig.~\ref{fig:direct dndz} for $N_{dn/dz} = 5$ (top panels) and
$N_{dn/dz}=10$ (bottom). The redshift bins cover the range $z = 0.6 - 1.6$
and have width $\Delta z = 0.2(0.1)$ for $N_{dn/dz}=5(10)$.

To quantify how well $f(z)$ (the normalized source distribution, as before)
can be reconstructed by the cross-correlation method, we treat its value in each bin as a free parameter, $f_i$,
in the Fisher matrix (imposing the normalization condition $\int dz \, f(z) = 1$).
For each $f(z)$ bin, we define one slice of spectroscopic galaxies with the same redshift range (thus giving rise
to $N_{dn/dz}$ spectroscopic slices). Each spectroscopic bin has a corresponding free galaxy bias parameter (as before).
Moreover, in each of these redshift bins, we allow for a free galaxy bias of the target ($p$) sample, and, in addition,
two free target bias parameters for the bins $z = 0 - 0.6$ and $z = 1.6 - \infty$.

Thus, in total, we consider $N_{dn/dz}$ $ss$ auto-spectra (the $s_i s_j$ cross-spectra are zero in the Limber approximation because of the absence
of redshift overlap), $N_{dn/dz}$ $sp$ cross-spectra, and one $pp$ auto-spectrum
as our (prospected) data set and $N_{\rm cosmo} + N_{b^{(s)}} + N_{b^{(p)}} + N_{dn/dz} = 7 + N_{dn/dz} + (N_{dn/dz} + 2) + N_{dn/dz}
= 9 + 3 N_{dn/dz}$ parameters (making 24 or 39 in practice).
We wish to stress again that the constraints on the target redshift distribution we will present
have any degeneracy with cosmological parameters taken into account and marginalized over.
No CMB prior is included so that any degeneracy between
$f(z)$ and cosmological parameters has to be broken by the galaxy spectra themselves.

Our main results are shown in Fig.~\ref{fig:direct dndz}. We first consider the case where the target galaxy bias $b^{(p)}(z)$
is known exactly (but the cosmological parameters and spectroscopic galaxy bias are marginalized over).
The resulting uncertainties in the reconstructed redshift distribution are indicated by the
black error bars in the top(bottom) panels for the case $N_{dn/dz} = 5(10)$ (for these errors, there is no difference between
the left and right panels). In both cases, errors as small as 2 \% can be reached on $f(z)$ in the
central bins. The uncertainties in $f(z)$ do not vary strongly between bins, but the relative uncertainties do because of the lower fiducial
values in the bins on the edge of the distribution. In those bins, the relative uncertainties are of order unity or larger.
The measurement of the redshift distribution corresponds to a determination of the sample's mean redshift
of $\sigma(\langle z \rangle) \approx 0.005$ (independent of the number of bins $N_{dn/dz}$).

The bounds above, however, rely strongly on our assumption that the galaxy bias is known exactly,
and will deteriorate when uncertainty in the bias (of the $p$ sample) is allowed.
In fact, if we allow the bias to be completely free (no prior) in the redshift bins discussed
above, the error bars on $f(z)$ approach infinity (not shown in the figure) and we are left with no information
on the redshift distribution. This was to be expected because of the exact degeneracy between a free $b^{(p)}(z)$
and $f(z)$.

We now consider the constraints when independent bias measurements and/or theoretical considerations allow us to place a
prior on the galaxy bias evolution, again employing the two types of priors discussed in Section \ref{sec:cc th}:
a diagonal prior $\sigma^{\rm bias}_{\rm diag}$, or a prior on the slope of the bias-redshift relation (assuming linearity),
$\sigma^{\rm bias}_{\rm lin}$.
The left panels show that a weak prior $\sigma^{\rm bias}_{\rm diag} = 0.1$ (i.e.~$\sigma(b^{(p)}_i) = 0.1/\sqrt{\Delta z}$, with $\Delta z$
the bin width) makes it possible again to measure $f(z)$, although still with rather large error bars (typical relative uncertainties
$\sim$ 30 \%, $\sigma(\langle z \rangle) \approx 0.02$).
Qualitatively similar results (but somewhat stronger) can be obtained by imposing $f(z)$
to be a linear function of redshift and applying a weak prior $\sigma^{\rm bias}_{\rm lin} = \sigma(b'^{(p)}) = 1$ (right panels).
Applying priors an order of magnitude better ($\sigma^{\rm bias}_{\rm diag}=0.01$ or $\sigma^{\rm bias}_{\rm lin}=0.1$,
see Figure \ref{fig:direct dndz})
is almost equivalent to knowing the bias exactly for the purpose of redshift distribution reconstruction.

We have shown above that to reconstruct a general galaxy redshift distribution using the cross-correlation technique,
it is crucial to have prior knowledge of the galaxy bias evolution. On the bright side,
the required bias prior for the method to be successful is not very strict and might be within reach (under the simple assumptions
made in this forecast). In the analysis with a cosmic shear focus in the previous sections, by contrast, we had found
that a bias prior, while useful, was not as important as here. In particular, even in the absence of
such a prior, information on the photo-$z$ parameters (and thus on $f(z)$) could be obtained
from the cross-correlation method. The crucial difference, however, in the former analysis
is that the true underlying distribution $dn/dz(z)$ was assumed to be known and that multiple source bins
were used. In that case, at the redshifts where the true redshift distributions
of neighboring tomographic slices overlap, we are sensitive to both $f_i(z) b^{(p)}(z)$ and $f_{i+1}(z) b^{(p)}(z)$,
where $f_i(z)$ and $f_{i+1}(z)$ are the normalized redshift distributions in neighboring bins.
The combination (specifically the ratio) of these quantities contains information on the
photo-$z$ parameters that does not suffer from the bias degeneracy. Therefore,
the photo-$z$ parameters could be constrained (weakly) even without making assumptions
about $b^{(p)}(z)$.
Note however that this is only the case because the galaxy bias was assumed to be a function
of redshift only. An additional dependence on color for instance would require us to model
the bias of galaxies in separate tomographic bins as independent functions, thus
reinstating the bias degeneracy.

\subsection{Lensing requirements on the redshift distribution measurement}
\label{subsec:lensing req}

We now isolate the main question on the other end of the procedure of using cross-correlations
to improve weak lensing as a cosmological probe: {\it what do we need to know about the lensing source
redshift distribution to optimize cosmological parameter constraints from cosmic shear?}
For simplicity, we will study this question for the case of a single source bin, with the same
fiducial redshift distribution (and parametrization in terms of redshift bins) as in the previous subsection.
Considering variations $\delta f(z)$ from the fiducial $f(z)$, the effects of certain modes on the shear power
spectrum will be orthogonal to the effects of cosmological parameters so that uncertainty in these modes
would not affect cosmological parameter constraints.
We here ask which modes/components of $f(z)$ we {\it do} need to know because they are degenerate with
cosmological parameters. Note that these conclusions will hold only for a given model
and might change if we include, e.g., massive neutrinos, etc.

We study this question by considering the cosmological parameter bias $\delta p$
induced by assuming the wrong source distribution, with $\delta f(z)$
the difference between the assumed and the true distribution.
Note that this bias is closely related to the increase in variance $\sigma^2(p)$
in the case where uncertainty in $f(z)$ is modeled properly and marginalized over
\footnote{Specifically, the requirement for the parameter bias $\delta p$ to be small compared to the uncertainty $\sigma(p)$
is equivalent to the requirement for the relative change in parameter uncertainty (due to marginalization
over uncertainty in $f(z)$) to be small compared to the uncertainty in $p$ in the case where $f(z)$ is known.}.

\begin{figure*}
  \begin{center}{
  \includegraphics[width=0.48\columnwidth]{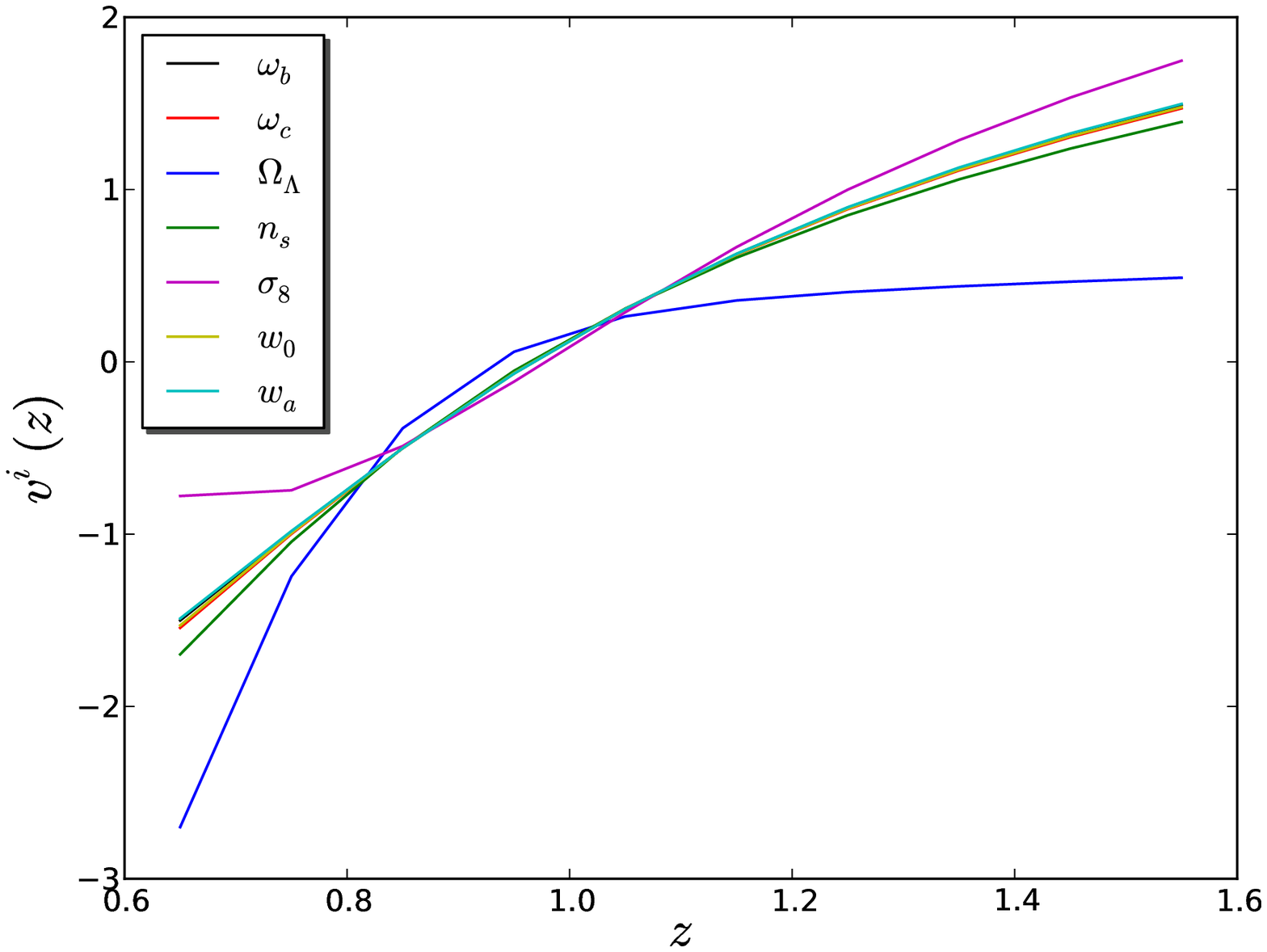}
  \includegraphics[width=0.48\columnwidth]{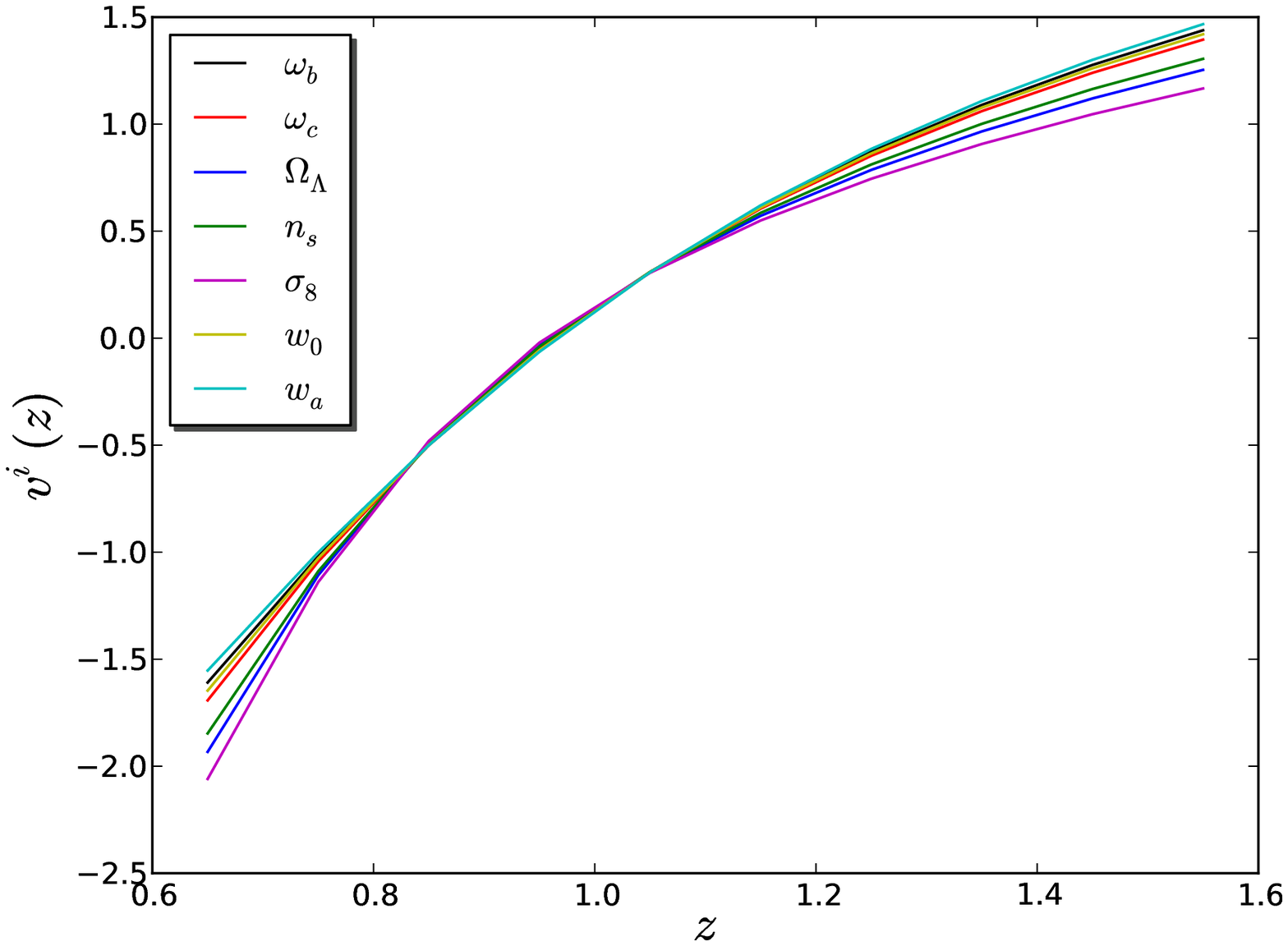}
  }
  \end{center}
  \caption{The modes $v^i(z)$ defining what type of misestimate of $f(z)$ would most bias each cosmological parameter
  if this offset is not modeled properly.
  The bias in a cosmological parameter is given by $\alpha_i \equiv \int dz \, v^i(z) \, \delta f(z)$, multiplied by
  the factor given in Tables VII and VIII. Alternatively, if freedom in $f(z)$
  is properly modeled, marginalizing over the possible variations in $f(z)$ causes additional cosmological parameter uncertainty
  given by the uncertainty in $\alpha_i$ multiplied by the factors given in the tables mentioned above.
  {\it Left:} results for the case of a HSC-like lensing survey (but using only one tomographic bin, see text)
  with a Planck prior. {\it Right:} Same as left, but with $f_{\rm sky} = 1$ for the lensing survey.}
  \label{fig:vi(z)}
\end{figure*}

In the Fisher matrix formalism, the parameter bias is given by
\beq
\label{eq:bias}
\delta p_i = - \sum_{j=1}^{N} \sum_{k=1}^{N_{dn/dz}} (F^{(N)})^{-1}_{ij} \, F^{(N+N_{dn/dz})}_{jk} \, \delta f_k \equiv \sum_{k=1}^{N_{dn/dz}} \frac{\pa \delta p_i}{\pa \delta f_k} \, \delta f_k,
\eeq
where $\delta p_i$ is the bias in the $i$-th cosmological parameter,
$N$ the number of parameters {\it not} including the $N_{dn/dz}$ parameters describing the source distribution,
$F^{(N_{\rm cosmo})}$ the Fisher matrix restricted to those parameters,
$F^{(N+N_{dn/dz})}$ the Fisher matrix for the complete parameter space, and $\delta f_k$ the offset in the binned values
of the normalized source redshift distribution $f(z)$.
In the limit of a large number of redshift bins, it is convenient to approximate this
in terms of continuous functions, and to write the parameter bias as the product of $\alpha_i$, which is the inner product of
$\delta f(z)$ with a mode picking out the redshift dependence degenerate with $p_i$, and a factor $\pa \delta p_i/\pa \alpha_i$, determining the amplitude
of the parameter bias. In equations,
\beq
\delta p_i =  \frac{\pa \delta p_i}{\pa \alpha_i} \, \alpha_i,
\eeq
with
\beq
\alpha_i \equiv \int dz \, v^i(z) \, \delta f(z),
\eeq
and the normalized mode defining the inner product (assuming uniform redshift bins $\Delta z_i = \Delta z$ for all bins $i$)
given by
\beq
v^i(z_k) \propto \frac{\pa \delta p_i}{\pa \delta f_k}, \quad {\rm s.t.} \quad \int dz \, \left( v^i(z) \right)^2 = 1
\eeq
(this fully defines $v^i(z)$, except for its sign, which is arbitrary).
An alternative interpretation of $\alpha_i$ is as the coefficient of the mode $v^i(z)$ in an expansion of $\delta f(z)$,
\beq
\delta f(z) = \alpha \, \hat{v}(z) + \sum_j c_j \, v_{\perp}^j(z),
\eeq
where the modes in the second term on the right hand side can be part of any basis with
\beq
\int dz \, v^j_{\perp}(z) \, \hat{v}(z) = 0.
\eeq

In Figure \ref{fig:vi(z)} (left panel), we show the modes $v^i(z)$ for the seven cosmological parameters considered in this work for the case
of cosmic shear data with a Planck prior. Strikingly, the mode $v^i(z)$ is almost the same for each parameter except $\Omega_{Lambda}$,
showing that the only property
of $\delta f(z)$ that matters is its inner product with this set of two distinct modes. Uncertainty in orthogonal components of $\delta f(z)$
would not lead to cosmological parameter bias (or additional uncertainty). The main reason that all these modes are so similar,
even though they describe the degeneracy directions with very different cosmological parameters,
is the inclusion of the CMB prior. This prior already constrains rather tightly a large number of parameters.
Considering the principal components of the CMB-only Fisher matrix, we find that weak lensing only moderately improves three of these
(and the other four not at all).
Thus, the only variations in $f(z)$ that can affect joint cosmological parameter constraints are the ones biasing
parameters in this three-dimensional subvolume of the total cosmological parameter space. This significantly narrows down the range of
possible modes. In fact, we have checked that,
when only the weak lensing data are considered, the $v^i(z)$ modes differ much more strongly, confirming that
the reason for them being the same in our case is the inclusion of the CMB prior.

Table \ref{tab:moderesponse} shows the parameter bias resulting from an offset in the coefficient $\alpha_i$ (which for a given $\delta f(z)$
hardly depends on $i$ because of the near-universality of the mode $v^i(z)$). The left column shows the change in parameter bias per unit
change in $\alpha_i$, and the right column shows the same quantity normalized by the uncertainty in the cosmological parameter.
Judging from the second column, the parameters most affected by a misestimate of (or by uncertainty in) $f(z)$ are $w_0$
and $w_a$. Specifically, we find $(\pa \delta w_a/\pa \alpha)/\sigma(w_a) =-6.4$. This means that if we can limit our misestimate
of the relevant mode of $f(z)$ to be
$\alpha_{w_a} < 1./6.4 = 0.16$, then the parameter bias will be small, $\delta w_a < \sigma(w_a)$ (note that errors are added in quadrature).
Equivalently, limiting the uncertainty $\sigma(\alpha_{w_a}) < 0.16$, means that the additional uncertainty in $w_a$
is small in case the $f(z)$ uncertainty is marginalized over. Since the relevant mode $v^i(z)$ is so similar for each parameter
(the only exception being $\Omega_{\rm Lambda}$, which has a weak response, $-0.20$, to variations in $v^{\Omega_{\Lambda}}(z)$),
and since $w_a$ is the most strongly affected parameter, the requirement for the other parameters to be negligibly affected
is less stringent than this.

\begin{table*}[hbt!]
\begin{center}
\small
\begin{tabular}{c|cc}
\hline\hline
  &  $\pa \delta p/\pa \alpha$ & $(\pa \delta p/\pa \alpha)/\sigma_0(p)$ \\
\hline\hline
$\omega_b$ & 0.00016 & 1.3 \\
$\omega_c$ & -0.0021 & -1.9 \\
$\Omega_\Lambda$ &  -0.033 & -0.20 \\
$n_s$ &  0.0019 & 0.58 \\
$\sigma_8$ &  0.092 & 0.51 \\
$w_0$ & 4.7 & 3.7 \\
$w_a$ & -16 & -6.4 \\
\hline\hline
\end{tabular}
\caption{{\it Left column:} Response of cosmological parameter bias to variations in the relevant component ($\alpha_i$, defined by the modes $v^i(z)$
shown in the left panel of Figure 6) of the offset between the assumed and true source redshift distributions.
An HSC-like survey, together with CMB information from Planck, is assumed.
{\it Right column:} Same as left, but normalized by the parameter uncertainty in the case of perfectly known source distribution.
}
\label{tab:moderesponse}
\end{center}
\end{table*}

In summary, in the simple case considered above of cosmic shear in a single tomographic bin, the requirement on
our knowledge of the source distribution is
\beq
\int dz \, \delta f(z) \, v^{w_a}(z) < 0.16,
\eeq
where $v^{w_a}(z)$ is given by the cyan curve in Figure \ref{fig:vi(z)}.
In the present case (single lensing source bin with CMB prior, etc.),
this constraint is more or less satisfied even when the source distribution is self-calibrated
using the lensing information (i.e.~no external information from cross correlations).
Specifically, we find $\sigma(\alpha_{w_a}) = 0.17$ (the uncertainties of the other coefficients $\sigma(\alpha_i)$ are in the range $0.16 - 2.5$ for all parameters,
but we have confirmed that $w_a$ is the parameter most affected by uncertainty in $dn/dz(z)$)
so that even without additional information, the uncertainty in $f(z)$ does not affect the WL+Planck cosmological constraints much.

However, upon further inspection, the reason for this is simply that a single source bin at $z_{\rm ph} = 0.8 - 1.2$,
with HSC-like survey specifications, does not add much information to the CMB-only case even with perfect knowledge of $f(z)$. Variations in $f(z)$
thus do not have a large effect on the final cosmological constraints and the requirement on the knowledge of $f(z)$ is weak.
We therefore consider next the more interesting case
where the lensing measurement {\it does} add significant information
to the Planck-only case. We achieve this by simply considering a full-sky shear measurement of a source sample
with the same properties as above (except for the sky coverage).
The resulting modes $v^i(z)$ are shown in the right panel of Figure \ref{fig:vi(z)} and the response of cosmological parameters
to variations in the coefficients $\alpha_i$ in Table \ref{tab:moderesponse2}.
The modes $v^i(z)$ are now even more similar across the set of cosmological parameters.
Since now the lensing contributes more to parameter constraints, they are more sensitive to
uncertainty in $f(z)$, as is shown best in the second column of Table \ref{tab:moderesponse2}.

The strictest requirement on $f(z)$ again comes from $w_a$. In order not to bias this parameter
significantly, $\alpha_{w_a} < 1./9.2 < 0.11$ is needed. The uncertainty in $\alpha_{w_a}$
from the shear power spectrum itself (+ Planck) is $\sigma(\alpha_{w_a}) = 0.14$ (in general,
$\sigma(\alpha_i) = 0.14 - 0.30$). Thus, marginalizing
over the uncertainty in $f(z)$, the constraint on $w_a$ is weakened by a factor $(1 + (9.2 \cdot 0.14)^2)^\ha \approx 1.6$
compared to the case of known $f(z)$ ($\sigma(w_a) = 2.3 \to 3.7$). Other parameter uncertainties increase by smaller factors,
but overall this is a significant degradation of cosmological information.
On the other hand, using the cross and auto spectra $sp+pp+ss$ (as used in the previous subsection,
1500 deg$^2$ sky coverage), fixing $b^{(p)}(z)$, we find a constraint $\sigma(\alpha_{w_a}) = 0.016$ (in fact, for all parameters,
$\sigma(\alpha_i) = 0.016$) so that the effect of $f(z)$ on parameter constraints becomes negligible.

\begin{table*}[hbt!]
\begin{center}
\small
\begin{tabular}{c|cc}
\hline\hline
  &  $\pa \delta p/\pa \alpha$ & $(\pa \delta p/\pa \alpha)/\sigma_0(p)$ \\
\hline\hline
$\omega_b$ & 0.00021 & 1.7 \\
$\omega_c$ & -0.0036 & -3.6 \\
$\Omega_\Lambda$ &  -0.47 & -4.4 \\
$n_s$ &  0.0058 & 1.9 \\
$\sigma_8$ &  -0.43 & -4.1 \\
$w_0$ & 7.7 & 7.6 \\
$w_a$ & -21 & -9.2 \\
\hline\hline
\end{tabular}
\caption{As Table VII, but with the lensing survey scaled up to cover the full sky. This makes the constraining power
of cosmic shear, as compared to that of the CMB, stronger, and therefore makes the effect of uncertainty in the lensing source distribution
more important.
}
\label{tab:moderesponse2}
\end{center}
\end{table*}

\subsection{Summary}

The study above has broken down the procedure followed in this paper into its two main components.
\begin{itemize}

\item

Weak lensing constraints are weakened if $f(z)$ is not known accurately enough, and we have quantified
above which properties of $f(z)$ need to be constrained, and with what precision. We have done this for the particular case
of a single source distribution (no tomography). The specific results will depend on many assumptions, but the methodology
above is of general application. Moreover, a result that appears robust against changes in fiducial source redshift distribution
is that the main property of $f(z)$ that affects cosmological constraints is an inner product of $f(z)$ with a mode
of the shape depicted in Figure \ref{fig:vi(z)} that crosses zero only once.
The dominant effect of such a mode is to shift the average redshift of the distribution.

\item

The other component, discussed in subsection \ref{subsec:ps general}, is how well the cross-correlation technique
can provide an external measurement of $f(z)$.  We have shown that this method can provide a strong measurement
of $f(z)$, provided that sufficient knowledge of the galaxy bias evolution $b^{(p)}(z)$ is available.
This measurement of $f(z)$ can then be propagated to a measurement of
the mode coefficients (inner products) $\alpha_i$,
which can be compared to the requirement of a cosmic shear cosmology analysis.
The quantitative results are strongly dependent on survey and sample assumptions,
but we have given an example for illustration. In general, it appears that with sufficient knowledge
of $b^{(p)}(z)$, the cross-correlation technique provides enough $f(z)$ information to restore
the power of cosmic shear to its level in the case of perfectly known $f(z)$, but the specific galaxy bias prior
requirement differs from case to case.

\end{itemize}

The analysis in this section takes a rather different approach than our main forecasts for realistic surveys in the previous
sections, but we hope that by isolating the phenomenology involved in the Fisher forecasts of those sections, we have provided a bit more
insight into those constraints.

\section{Discussion and Conclusions}
\label{sec:disc}

We have studied the use of cross-correlations between lensing source galaxies and an overlapping sample
of spectroscopic galaxies as a method to measure the source galaxy redshift distribution
and to thus improve cosmic shear as a cosmological probe.
We used the Fisher matrix formalism to for the first time directly forecast the impact on cosmological constraints
of this cross-correlation technique, focusing on dark energy constraints from two types of future experiments:
a ground based SuMIRe-like survey (HSC lensing + PFS redshift survey) and a space based EUCLID-like survey.
For the main results of this paper, we have considered the scenario where
the source galaxies have photometric redshifts, which are used to divide the source sample
into tomographic bins, so that the redshift distribution in each bin is known perfectly
(only) in the limit where the photo-$z$ distribution is known exactly.

We first considered weak lensing constraints in the absence of galaxy density cross-correlation
information and have shown that cosmic shear can strongly improve dark energy constraints
relative to the case of (unlensed) CMB information only
(increasing the dark energy figure of merit by factors of 20-300 for HSC-EUCLID),
if and only if
the photo-$z$ parameters (defining the photo-$z$ distribution) are known well.
We confirmed the well known result from the literature that
in order for weak lensing to reach its full potential as a dark energy probe,
the photo-$z$ distribution needs to be known at the level $\sigma(\sigma_z), \sigma(b_z) < 0.01$.

We then considered to what extent the cross-correlation technique can restore the cosmology constraints
from weak lensing by measuring the photo-$z$ parameters (we remind the reader that we do {\it not}
use information on cosmological parameters present in the galaxy cross- and auto-spectra, only the information
on the source redshift distribution). We list some key results below.

\begin{itemize}

\item

Starting with the case where there is no prior knowledge of the photo-$z$ parameters,
the cross-correlation technique results in strong improvements in
the forecasted weak lensing uncertainties.
For the SuMIRe-like survey, the effect of the cross-correlation information on the dark energy FOM
is equivalent to placing a prior $\sigma_{\rm prior} = \sigma(\sigma_z) = \sigma(b_z) \approx 0.04 - 0.05$
({\it known galaxy bias} - {\it free galaxy bias})
on the photo-$z$ parameters at all redshifts. For the EUCLID-like survey,
using the cross-correlations is equivalent to an even better known photo-$z$ distribution,
$\sigma_{\rm prior} \approx 0.005 - 0.010$. One reason for the increased success of the method in the
case of EUCLID is the fact that EUCLID's spectroscopic survey has much better coverage of the low redshift end of
the distribution.

\item

In the more realistic case where some level of prior knowledge of the photo-$z$ distribution is assumed,
e.g.~coming from calibration of the photo-$z$ estimator using galaxy spectra for a representative subsample of the source galaxies,
the cross-correlation approach still strongly improves constraints, unless the prior on the photo-$z$
distribution is very strong.
For example, for SuMIRe, with a prior $\sigma_{\rm prior} = 0.05$ on the photo-$z$ parameters,
including the information from the cross-correlation technique improves the dark energy FOM
by more than a factor $4 - 3$ ({\it known galaxy bias} - {\it free galaxy bias}) relative to the case without this information. For EUCLID, with the same photo-$z$ prior, the gains are even more spectacular,
giving a factor $40 - 17$ improvement. Only when the photo-$z$ calibration is better than $\sigma_{\rm prior} \approx 0.01 (0.002)$
for SuMIRe (EUCLID), do the benefits from the cross-correlation method become negligible ($\lesssim 10 \%$).
We do note that, even in the cases where the dark energy figure of merit is significantly enhanced by use of the cross-correlation information,
it does not reach all the way to the value that could have been obtained if the source redshift distribution were known perfectly.

\item

The power of the cross-correlation method, however, depends strongly on the assumed knowledge of the galaxy bias evolution
of the source sample.
We have modeled the galaxy bias as a free function of redshift, defined by independent bias values in a large number of redshift bins
and have considered both the case of {\it a priori} completely unknown values of these bias parameters
and various levels of prior knowledge (including knowing the galaxy bias exactly).
For example, again in the case with a photo-$z$ calibration at the
$\sigma_{\rm prior} = 0.05$ level, assuming exact knowledge of $b^{(p)}(z)$ yields a
$49 \% (66 \%)$ larger dark energy FOM (and therefore effective survey volume)
for SuMIRe (EUCLID)
than when no prior knowledge of $b^{(p)}(z)$ is assumed.
The optimal constraints of the {\it known bias} case can be approached
by imposing a bias prior $\sigma^{\rm bias}_{\rm diag} \lesssim 0.02$
($\sigma^{\rm bias}_{\rm diag} \lesssim 0.005$) for SuMIRe (EUCLID).
This prior can be seen as the prior on the galaxy bias per redshift bin of width $\Delta z = 1$,
see Section \ref{subsec:biasevol}.
As discussed in Section \ref{subsec:ps general}, the reason that the cross-correlation technique
still provides some information in the absence of a galaxy bias prior
can be explained by our simple model for the source distribution, which allows the
extraction of information from the overlap in tomographic bins
that does not suffer from the bias degeneracy. This would likely not work in practice however.

\end{itemize}

To gain more insight in the results described above,
we have included a section (Section \ref{direct})
showing to what degree a single redshift distribution
can be reconstructed in redshift bins using the cross-correlation method, in the absence
of photo-$z$ information.
In this case, we have confirmed that without any galaxy bias prior,
no information on the sample's redshift distribution can be obtained.
We have found that a reasonable reconstruction of the distribution
($\sim 30 \%$ uncertainties)
can be achieved with a bias prior $\sigma^{\rm bias}_{\rm diag} \approx 0.1$.
A prior an order of magnitude smaller results in an optimal (in the sense that it can not be improved by tightening the prior even more) reconstruction
of the redshift distribution, with error bars in individual redshift bins as small as $2 \%$.

We have in the same section determined, for each cosmological parameter, which component (or {\it mode}) of the source redshift distribution
needs to be known in order to not bias that parameter in a cosmic shear study.
We have demonstrated that, when weak lensing is combined with the CMB,
this mode varies very little between different cosmological parameters and predominantly describes a shift in the average redshift of the distribution.
With only weak lensing data (including the CMB prior, as always), the coefficient of this mode is typically not well measured,
leading to a degradation of cosmological constraints. However, the galaxy cross- (and auto-)spectra are capable of measuring this coefficient
much more accurately, thus explaining how the cross-correlation technique aids weak lensing as a cosmological probe.
\\

In summary, our results confirm that using cross-correlations to constrain the source redshift distribution
(whether on its own or, more realistically, in combination with photometric redshifts)
has the potential to significantly improve the constraining power of upcoming weak lensing
surveys, although the level of success depends strongly on our ability to constrain the bias evolution
of the source galaxies.
While these are very encouraging results, we have made several simplifications and additional research
is needed to clarify how the method is affected by changes in these assumptions.
For example, it would be interesting to go beyond the Gaussian description
of the photo-$z$ distribution and include, among other things,
outliers in the distribution. It would also be useful to study extensions of the cosmological
model considered in this work, including massive neutrinos, modifications of gravity, etc. Moreover,
it is not clear what the role will be of magnification bias (whether it will weaken the cross-correlation approach or improve it by
helping break the degeneracy between redshift distribution and galaxy bias).
Finally, the fact that the photo-$z$ parameters could be measured even in the absence of
a galaxy bias prior really hinges on the assumption that the galaxy bias is a function of redshift only.
A more general treatment would allow for dependence on galaxy properties such as color (so that the bias
of galaxies in different tomographic bins at the same true redshift is not necessarily equal),
which, if left otherwise unconstrained by additional data or modeling, would worsen the constraints from
the cross-correlation method.

\section{acknowledgments}

The authors thank Carlos Cunha, Patrick MacDonald, Jeffrey Newman, David Schlegel, David Spergel and Masahiro Takada for useful discsussions.
Part of the research described in this paper was carried out at the Jet Propulsion Laboratory, California Institute of Technology,
under a contract with the National Aeronautics and Space Administration. This work is supported by NASA ATP grant 11-ATP-090.

\bibliography{refs}

\end{document}